\documentclass[aps, prd, amsmath, floats, floatfix, twocolumn, nofootinbib, superscriptaddress]{revtex4-1}
\usepackage[pdftex]{graphicx}
\usepackage{color}
\usepackage{latexsym}
\usepackage[colorlinks]{hyperref}

\newcommand{\be}{\begin{eqnarray}}
\newcommand{\ee}{\end{eqnarray}}
\newcommand{\rar}{\rightarrow}

\begin{document}

\title{Constraints on the spacetime metric around seven ``bare'' AGNs\\using X-ray reflection spectroscopy}

\author{Ashutosh~Tripathi}
\affiliation{Center for Field Theory and Particle Physics and Department of Physics, Fudan University, 200438 Shanghai, China}

\author{Jinli~Yan}
\affiliation{Center for Field Theory and Particle Physics and Department of Physics, Fudan University, 200438 Shanghai, China}

\author{Yuchan~Yang}
\affiliation{Center for Field Theory and Particle Physics and Department of Physics, Fudan University, 200438 Shanghai, China}

\author{Yunfeng~Yan}
\affiliation{Center for Field Theory and Particle Physics and Department of Physics, Fudan University, 200438 Shanghai, China}

\author{Marcus~Garnham}
\affiliation{Department of Physics \& Astronomy, University of Nottingham, University Park, Nottingham NG7 2RD, UK}
\affiliation{Center for Field Theory and Particle Physics and Department of Physics, Fudan University, 200438 Shanghai, China}

\author{Yu~Yao}
\affiliation{Center for Field Theory and Particle Physics and Department of Physics, Fudan University, 200438 Shanghai, China}

\author{Songcheng~Li}
\affiliation{Center for Field Theory and Particle Physics and Department of Physics, Fudan University, 200438 Shanghai, China}

\author{Ziyu~Ding}
\affiliation{Center for Field Theory and Particle Physics and Department of Physics, Fudan University, 200438 Shanghai, China}

\author{Askar~B.~Abdikamalov}
\affiliation{Center for Field Theory and Particle Physics and Department of Physics, Fudan University, 200438 Shanghai, China}

\author{Dimitry~Ayzenberg}
\affiliation{Center for Field Theory and Particle Physics and Department of Physics, Fudan University, 200438 Shanghai, China}

\author{Cosimo~Bambi}
\email[Corresponding author: ]{bambi@fudan.edu.cn}
\affiliation{Center for Field Theory and Particle Physics and Department of Physics, Fudan University, 200438 Shanghai, China}

\author{Thomas~Dauser}
\affiliation{Remeis Observatory \& ECAP, Universit\"{a}t Erlangen-N\"{u}rnberg, 96049 Bamberg, Germany}

\author{Javier~A.~Garc{\'\i}a}
\affiliation{Cahill Center for Astronomy and Astrophysics, California Institute of Technology, Pasadena, CA 91125, USA}
\affiliation{Remeis Observatory \& ECAP, Universit\"{a}t Erlangen-N\"{u}rnberg, 96049 Bamberg, Germany}

\author{Jiachen~Jiang}
\affiliation{Institute of Astronomy, University of Cambridge, Cambridge CB3 0HA, UK}

\author{Sourabh~Nampalliwar}
\affiliation{Theoretical Astrophysics, Eberhard-Karls Universit\"at T\"ubingen, 72076 T\"ubingen, Germany}

\begin{abstract}
We present the study of a sample of seven ``bare'' active galactic nuclei (AGN) observed with \textsl{Suzaku}. We interpret the spectrum of these sources with a relativistic reflection component and we employ our model {\sc relxill\_nk} to test the Kerr nature of their supermassive black holes. We constrain the Johannsen deformation parameters $\alpha_{13}$ and $\alpha_{22}$, in which the Kerr metric is recovered when $\alpha_{13} = \alpha_{22} = 0$. All our measurements are consistent with the hypothesis that the spacetime geometry around these supermassive objects is described by the Kerr solution. For some sources, we obtain quite strong constraints on $\alpha_{13}$ and $\alpha_{22}$ when compared to those found in our previous studies. We discuss the systematic uncertainties in our tests and the implications of our results. 
\end{abstract}

\maketitle


\section{Introduction}

Einstein's theory of general relativity is one of the pillars of modern physics. It has successfully passed a large number of observational tests~\cite{will}. However, the strong field regime is still largely unexplored. The best laboratory for testing strong gravity is the spacetime around astrophysical black holes~\cite{r1,r2,r3,r4,r5,eck}. In 4-dimensional general relativity, the only stationary and asymptotically flat vacuum black hole solution, which is regular on and outside the event horizon, is described by the Kerr metric~\cite{h1,h2,h3}. The spacetime geometry around astrophysical black holes formed from gravitational collapse is thought to be well approximated by this solution~\cite{b1,b2}. However, macroscopic deviations from the Kerr metric may be possible if general relativity is not the correct theory of gravity, as well as in the presence of macroscopic quantum gravity effects~\cite{n1,n2,n3,n4}.

X-ray reflection spectroscopy refers to the study of the reflection spectrum of accretion disks around black holes~\cite{k1,k2}. The temperature of the inner part of a thin accretion disk is in the soft X-ray band (0.1-1~keV) for stellar-mass black holes of $\sim 10$~$M_\odot$ and in the optical/UV band (1-100~eV) for supermassive black holes of $\sim 10^6$-$10^9$~$M_\odot$. Thermal photons from the accretion disk can have inverse Compton scattering off free electrons in the corona, which is a hot ($\sim 100$~keV), usually compact and optically thin, medium close to the black hole. For instance, the corona may be the base of a jet or the atmosphere above the accretion disk. The inverse Compton scattering generates a power-law spectrum. Comptonized photons can illuminate the disk, producing a reflection component with some emission lines. The most prominent features of the reflection spectrum are usually the iron K$\alpha$ complex around 6~keV (depending on the ionization of iron nuclei) and the Compton hump around 20~keV.

The reflection spectrum at the emission point in the rest-frame of the gas in the accretion disk is determined by atomic physics. These photons then propagate in the strong gravitational field of the black hole and experience a number of relativistic effects (Doppler boosting, gravitational redshift, and light bending) before being detected by our instruments. In the presence of the correct astrophysical model, an accurate measurement of the reflection spectrum of the accretion disk can provide information about the spacetime metric in the strong gravity region and thus test the Kerr nature of astrophysical black holes~\cite{nk1,nk2,nk3,nk4,nk5,nk6,nk7,nk8}.

Recently, we have developed the relativistic reflection model {\sc relxill\_nk} to test the Kerr black hole hypothesis~\cite{noi0}. {\sc relxill\_nk} is the natural extension of the {\sc relxill} package~\cite{ref1,ref2} to non-Kerr spacetimes. The reflection spectrum at the emission point in the rest-frame of the gas in the accretion disk is processed by a convolution model for a parametric black hole spacetime. The background metric has a number of ``deformation parameters'' introduced to quantify possible deviations from the Kerr solution. This metric reduces to the Kerr metric when all the deformation parameters vanish, and describes a deformed spacetime when at least one of the deformation parameters is different from zero. From the comparisons of the theoretical predictions of {\sc relxill\_nk} with X-ray data of astrophysical black holes, we can constrain the value of these deformation parameters and test the Kerr nature of astrophysical black holes.

In the past year, we have applied our reflection model {\sc relxill\_nk} to a few stellar-mass and supermassive black holes, obtaining the first constraints on possible deviations from the Kerr geometry in the strong gravity region around these objects (see Ref.~\cite{shenzhen} for a summary). Constraints on some deformation parameters have been obtained from the analysis of \textsl{XMM-Newton}, \textsl{NuSTAR}, and \textsl{Swift} data of the supermassive black hole in 1H0707--495~\cite{noi1,noi1b}, from \textsl{Suzaku} data of the supermassive black holes in Ark~564~\cite{noi2} and Mrk~335~\cite{noi5}, and from \textsl{RXTE} data of the stellar-mass black hole in GX~339--4~\cite{noi3}. The most stringent constraints have been obtained from the analysis of a \textsl{NuSTAR} observation of the stellar-mass black hole in GS~1354--645~\cite{noi4} and from the combined analysis of \textsl{XMM-Newton} and \textsl{NuSTAR} data of the supermassive black hole in MCG--6--30--15~\cite{noi6}. All our measurements are consistent with the hypothesis that the spacetime metric around these objects is described by the Kerr solution. While our model has a number of systematic uncertainties, a fine cancellation among very different contributions sounds unlikely.

Motivated by our previous results, we want to try to further boost the limit of these constraints. In this paper, we present the study of a sample of seven ``bare'' active galactic nuclei (AGN), where by bare we mean that these sources have little or no complicating intrinsic absorption. These seven objects belong to the sample of the 25~bare AGN observed with \textsl{Suzaku} and studied in~\cite{w13}. We have selected the sources more suitable for our tests, because the sample of Ref.~\cite{w13} also includes observations with low quality data that do not permit one to constrain the metric of the black hole. The spectrum of these seven sources is simple, often they only show the power law from the corona and the reflection spectrum from the disk, so the number of free parameters is relatively low, reducing the parameter degeneracy and favoring the possibility of getting stronger constraints on the deformation parameters. The inner edge of the accretion disk of these sources is also very close to the compact object, and this is another ingredient helping to place stronger constraints on the deformation parameters, as we can probe the region very close to the object. Eventually, we find that all our measurements are consistent with the Kerr metric, confirming our previous results as well as the validity of the technique employed in our study. For some sources, we can obtain impressive constraints on possible deviations from Kerr, comparable to those obtained from GS~1354--645 in Ref.~\cite{noi4} and MCG--6--30--15 in Ref.~\cite{noi6}. We discuss these results and we point out that it would be interesting to repeat the analysis with different deformation parameters.

Our manuscript is organized as follows. In Section~\ref{s-metric}, we briefly review the parametric black hole background employed in our analysis. In Section~\ref{s-red}, we describe our data reduction. Section~\ref{s-ana} shows the spectral analysis and the constraints on the deformation parameters for every source in the sample. Our results are discussed in Section~\ref{s-dis}. Throughout the paper, we adopt the convention $G_{\rm N} = c = 1$ and a metric with signature $(-+++)$.


\section{Parametric black hole spacetimes \label{s-metric}}

Black hole tests with electromagnetic techniques can follow two possible strategies. In the literature, they are often referred to as, respectively, {\it top-down} and {\it bottom-up} approaches.

The top-down method is the most logical and natural one: we want to test general relativity against some specific theory of gravity in which uncharged black holes are not described by the Kerr solution. Observational data are thus analyzed with the Kerr model of general relativity and with the non-Kerr model of the other gravity theory to check which one can better explain the data. Such an approach presents two problems. First, there are a large number of theories of gravity beyond general relativity and none of them seems to be more motivated than the others, so we should repeat the test for every known gravity theory. Second, only in a very limited number of gravity theories do we know the rotating black hole solutions. In many gravity theories we only know the non-rotating black hole solutions, and for some theories we also know the solutions in the slow-rotation limit. On the contrary, we know that astrophysical bodies have a non-vanishing angular momentum and that black hole tests require very fast-rotating compact objects, while slow-rotating objects usually can only provide very poor constraints.

In the bottom-up approach, we employ a phenomenological parametrization of the spacetime that ideally could be used to describe the Kerr solution as well as any black hole solution in any alternative theory of gravity. The metric is characterized by the black hole mass, the black hole spin angular momentum, and by a number of deformation parameters that quantify possible deviations from the Kerr geometry. From the comparison of theoretical predictions with observations, we want to measure these deformation parameters and check whether their value is consistent with zero as required by the Kerr metric.

In this work, we will follow the bottom-up approach, which is more often employed in current black hole tests with electromagnetic techniques. While some parametric black hole spacetimes can reduce to black hole solutions of specific theories of gravity for suitable choices of a ``small'' number of deformation parameters, at the moment it is impossible to measure several deformation parameters at the same time due to a few different reasons, including calculation capabilities, parameter degeneracy, and quality of data. In such a case, one may question the validity of a similar approach and of the final measurements, because these are (at best) approximated solutions, so the accuracy on the deformation parameters is not under control. However, the spirit with which we employ these phenomenological metrics is not to measure the parameters of the system but rather to perform a null experiment, in which we expect to recover the Kerr metric but we want to check this null hypothesis. Eventually the goal is to test several possible deformations from the Kerr geometry and try to obtain stronger and stronger constraints. In the case of a clear non-vanishing measurement of a certain deformation parameter, the strategy should change in order to investigate the right form of the spacetime metric around the black hole.

As the parametric black hole background, we choose the Johannsen metric with the deformation parameters $\alpha_{13}$ and $\alpha_{22}$~\cite{tj}, because these are the two deformation parameters with the strongest impact on the reflection spectrum~\cite{noi0}. In Boyer-Lindquist-like coordinates, the line element reads
\be\label{eq-jm}
ds^2 &=&-\frac{\Sigma\left(\Delta-a^2A_2^2\sin^2\theta\right)}{B^2}dt^2
+\frac{\Sigma}{\Delta}dr^2+\Sigma d\theta^2 \nonumber\\
&&-\frac{2a\left[\left(r^2+a^2\right)A_1A_2-\Delta\right]\Sigma\sin^2\theta}{B^2}dtd\phi \nonumber\\
&&+\frac{\left[\left(r^2+a^2\right)^2A_1^2-a^2\Delta\sin^2\theta\right]\Sigma\sin^2\theta}{B^2}d\phi^2 \, ,
\ee
where $M$ is the black hole mass, $a = J/M$, $J$ is the black hole spin angular momentum, and
\be
&&  \Sigma = r^2 + a^2 \cos^2\theta \, , \qquad
\Delta = r^2 - 2 M r + a^2 \, , \nonumber\\
&& B = \left(r^2+a^2\right)A_1-a^2A_2\sin^2\theta \, .
\ee
The functions $A_1$ and $A_2$ are defined as
\be
A_1 = 1 + \alpha_{13}\left(\frac{M}{r}\right)^3 \, , \quad
A_2 = 1 + \alpha_{22}\left(\frac{M}{r}\right)^2 \, .
\ee
$\alpha_{13}$ and $\alpha_{22}$ are the two deformation parameters and the Kerr solution is recovered when $\alpha_{13} = \alpha_{22} = 0$. In this paper, we will only consider the possibility that one of the deformation parameters is non-vanishing; that is, we will consider two cases, one in which we want to measure $\alpha_{13}$ assuming $\alpha_{22} = 0$ and one in which we want to determine $\alpha_{22}$ assuming $\alpha_{13} = 0$. This is because our current version of {\sc relxill\_nk} can only work with one free deformation parameter at a time. Note that we do not assume that $\alpha_{13}$ and $\alpha_{22}$ are some small metric perturbations and that there are some higher order terms in $M/r$. We want to perform a null experiment, so we introduce some parameters to quantify possible deviations from the Kerr background and we want to check whether their measurements are consistent with zero.

In our analysis, we will ignore spacetimes with pathological properties (spacetime singularities, regions with closed time-like curves, etc.). This requires some restrictions on the possible values of the spin parameter $a_* = a/M = J/M^2$ and of the deformation parameters $\alpha_{13}$ and $\alpha_{22}$. As for the Kerr metric, we require
\be
- 1 < a_* < 1 \, ,
\ee 
which is the condition for the existence of the event horizon (for $| a_* | > 1$ there is no horizon and the central singularity is naked). For the deformation parameters $\alpha_{13}$ and $\alpha_{22}$, we have to impose (see Refs.~\cite{tj,noi3} for the details).
\be
\label{eq-constraints}
&& \alpha_{13} > - \frac{1}{2} \left( 1 + \sqrt{1 - a^2_*} \right)^4 \, , 
\nonumber\\
&& - \left(1 + \sqrt{1 - a_*^2} \right)^2 < \alpha_{22} < \frac{\left( 1 + \sqrt{1 - a^2_*} \right)^4}{a_*^2} \, .
\ee

Note that deviations from the Kerr metric due to ``standard'' physics are typically extremely small and completely negligible for our tests. The spacetime metric should quickly reduce to the Kerr solution after the formation of the black hole due to the emission of gravitational waves~\citep{price72}. The impact of the mass of the accretion disk, or of nearby stars, is completely negligible~\citep{b2,k4,berlin}. The equilibrium electric charge can be reached very soon because of the highly ionized host environment and its value is extremely small for macroscopic bodies~\citep{b1}. The impact of all these effects have been estimated in previous studies and turn out to be completely negligible as long as we do not aim at measuring the spin or the deformation parameters with a precision of 6~digits or better, which is well beyond the capabilities of current theoretical models and of the quality of current data.


\begin{table*}
 \centering
 \caption{List of the seven sources selected for our study and the corresponding \textsl{Suzaku} observations. Redshifts are taken from NASA Extragalactic Database (NED). The Galactic column density is calculated from~\cite{nH}.}
\vspace{0.3cm}
\begin{tabular}{ccccccc}
Source & \hspace{0.0cm} $z$ \hspace{0.0cm} & \hspace{0.0cm} $N_{\rm H,Gal}$ \hspace{0.0cm} & \hspace{0.0cm} Observation ID \hspace{0.0cm} & \hspace{0.0cm} Observation Date \hspace{0.0cm} & \hspace{0.0cm} Exposure (ks) \hspace{0.0cm} & Cts/sec \\ 
\hline
Ton~S180 &0.062&1.36&701021010&09/12/2006&108&$0.818\pm0.003$\\
RBS~1124 &0.208&1.52&702114010&14/04/2007&79&$0.229\pm0.001$\\
\hspace{0.0cm} Swift~J0501.9--3239 \hspace{0.0cm} &0.012&1.84&703014010&11/04/2008&36&$2.408\pm0.006$\\
Ark~120 &0.033&14.5&702014010&01/04/2007&91&$2.256\pm0.004$\\
1H0419--577 &0.104&1.34&702041010&25/07/2007&179&$1.453\pm0.002$\\
PKS~0558--504&0.137&3.9&701011010&17/01/2007&20&$1.401\pm0.006$\\
&&&701011020&18/01/2007&19&$1.987\pm0.007$\\
&&&701011030&19/01/2007&21&$1.599\pm0.006$\\
&&&701011040&20/01/2007&20&$2.038\pm0.007$\\
&&&701011050&21/01/2007&20&$2.067\pm0.007$\\
Fairall~9 &0.047&3.43&702043010&07/06/2007&145&$1.971\pm0.003$\\
\end{tabular}
\vspace{0.3cm}
\label{tab1}
\end{table*}

\section{Observations and Data Reduction \label{s-red}}

Our starting point is the list of the 25~bare AGN observed with \textsl{Suzaku} and studied in~\cite{w13}. These sources have typically a simple spectrum, with little or no intrinsic absorption complicating their spectral analysis. According to the analysis in~\cite{w13} in which the Kerr metric is assumed, the spin parameter of these supermassive black holes is typically very high; that is, the inner edge of the accretion disk is very close to the compact object and the radiation emitted by the inner part of the disk is strongly affected by the properties of the strong gravity region. These are all useful properties for our tests, as we want to reduce the systematic uncertainties related to the astrophysical model and we want to have spectra significantly affected by the properties of the background metric. From this list, we have already studied Ark~564 in~\cite{noi2} and Mrk~335 in~\cite{noi5}. We select seven more sources. We choose those with the simplest spectrum (according to the analysis in~\cite{w13}), with no intrinsic absorption, high value of the spin parameter measured in~\cite{w13}, and with good \textsl{Suzaku} data. The list of the seven sources selected for our study is shown in Tab.~\ref{tab1}.

\textsl{Suzaku} had four X-ray imaging spectrometer (XIS) CCD detectors. Three of the detectors were front-illuminated and one was back-illuminated~\cite{koyama}. In this paper, we consider only front-illuminated CCDs because of their larger effective area around 6~keV and lower background at higher energies as compared to the back-illuminated CCD. One of the front-illuminated chips, XIS2, experienced charge leakage on 6 November 2007 and observations after that date do not have XIS2 data. We use {\sc heasoft} v6.24 and CALDB version 20180312 for data reduction. Raw data are used to extract filtered events using AEPIPELINE. These events are then read into XSELECT for extracting the spectrum~\cite{koyama}. The event files are first screened using standard criterion for XIS given in SUZAKU ABC guide and good time intervals (GTIs) are generated. The size of source regions varies from source to source but typically has a radius around 3.5~arcmin. Background regions of the same size are taken far from the source and avoid corners of the chip in order to exclude any contamination. The response files are generated using the heasoft script XISRMFGEN and ancillary files are obtained using the script XISSIMARFGEN assuming point-like sources. The spectra and response files of all available front-illuminated chips, XIS0, XIS2 (if applicable), and XIS3, are combined using the {\sc FTOOL} ADDASCASPEC. Lastly, the combined spectra are rebinned into 50~photons per bin in order to apply $\chi^2$ statistics. We do not use the data in the energy range 1.7-2.5~keV because of calibration uncertainties\footnote{https://heasarc.gsfc.nasa.gov/docs/suzaku/analysis/sical.html}.

For PKS~0558--504, we have five observations taken at different times. Since we are interested in average spectral properties of the source and the individual spectra are similar, we can combine these observations together.

In what follows, we do not show the results of the spectral analysis that includes the HXD PIN data because their quality is poor and does not permit us to improve the fits. Moreover, the aim of the present work is to present a survey of several sources and to show the potentialities of X-ray reflection spectroscopy for testing the Kerr nature of astrophysical black holes. We leave to a future work a more detailed analysis of every source including data from other X-ray missions.


\section{Spectral analysis \label{s-ana}}

For the spectral analysis, we use XSPEC v12.9.1~\cite{arnaud}. We employ the following models: {\sc tbabs}, {\sc zpowrlw}, {\sc xillver}, {\sc relxill\_nk}, and {\sc zgauss}.

{\sc tbabs} takes the Galactic absorption into account~\cite{wilms}; the hydrogen column density is frozen to the value calculated using~\cite{nH,nH2}. The value for every source is reported in Tab.~\ref{tab1}.

{\sc zpowrlw} describes the power-law spectrum with exponential cut-off of the corona and has four parameters: the photon index $\Gamma$, the cut-off energy $E_{\rm cut}$, the redshift of the source $z$, and the normalization. In our fits, $\Gamma$ is always free, $z$ is frozen to the value of the source, and the normalization is free. $E_{\rm cut}$ is frozen to 300~keV for all sources as our data end at 10~keV and we are unable to constrain this parameter. While $E_{\rm cut}$ may have an effect in the reflection model in the soft energy band~\cite{g15}, this cannot be seen with our \textsl{Suzaku} data because of the lack of a sufficiently high signal to noise ratio.

{\sc xillver} can describe a non-relativistic reflection component from possible cold material at large distance~\cite{ref3,ref4}. {\sc relxill\_nk} is employed to describe the relativistic reflection component from the accretion disk~\cite{noi0}. When the data require both {\sc xillver} and {\sc relxill\_nk}, their common parameters are tied with the exception of the ionization: in {\sc xillver} $\log\xi$ is frozen to 0 and in {\sc relxill\_nk} $\log\xi$ is always free. $\Gamma$ and $E_{\rm cut}$ are tied to the values of the parameters in {\sc zpowrlw}, but the reflection fraction in {\sc xillver} and {\sc relxill\_nk} is set to $-1$ because we already have {\sc zpowrlw}. The emissivity profile of the disk is described by a broken power law with three parameters: inner emissivity index $q_{\rm in}$, outer emissivity index $q_{\rm out}$, and breaking radius $R_{\rm br}$. In our analysis, for some sources we leave the three parameters free, for some sources we leave $q_{\rm in}$ and $R_{\rm br}$ free and we impose $q_{\rm out} = 3$ (Newtonian limit at large radii for a lamppost corona), and in other cases we describe the emissivity with a simple power law, so we have $q_{\rm in}$ free and $q_{\rm out} = q_{\rm in}$. The choice is determined by the quality of the data and of the fit for every source. The reflection spectrum calculated by {\sc relxill\_nk} also depends on the inclination angle of the disk with respect to the line of sight of the observer, $i$, the spin parameter of the black hole, $a_*$, and the deformation parameter, either $\alpha_{13}$ or $\alpha_{22}$. The inner edge of the accretion disk is always assumed to be at the innermost stable circular orbit (ISCO) of the spacetime, while the outer radius is frozen at 400~gravitational radii, which is large enough that its exact value is irrelevant as at large radii the emissivity is weaker and weaker.

AGN often show narrow emission lines consistent with highly ionized iron lines: Fe~XXV (6.67 keV) and Fe~XXVI (6.97 keV). We model these lines with {\sc zgauss} and we set the line width to 10~eV (the exact value is not important because it is much smaller than the energy resolution of the instruments).

\begin{table*}
 \centering
 \caption{$\chi^2/\nu$ for a model described by a power law ({\sc pow}), a power law with a relativistic reflection component ({\sc pow + rel}), and a power law with both a relativistic and a non-relativistic reflection component ({\sc pow + rel + xill}). In the second and third rows, $\alpha_{13}$ is free and $\alpha_{22} = 0$. In the fourth and fifth rows, $\alpha_{13} = 0$ and $\alpha_{22}$ is free.}
\vspace{0.2cm} 
\begin{tabular}{cccccccc}
Model & Ton~S180 $\,$ & RBS~1124 $\,$ & Ark~120 $\,$ & Swift~J0501.9--3239 $\,$ & 1H0419--577 $\,$ & PKS 0558--504 $\,$ & Fairall~9 \\
\hline
{\sc pow} & 2.157 & 1.041 & 4.076 & 1.542 & 1.410 & 2.081 & 3.026 \\
\hline
{\sc pow + rel} ($\alpha_{13}$) & 1.030 & 0.931 & 1.302 & 1.092 & 1.070 & 1.056 & 1.336 \\
{\sc pow + rel + xill} ($\alpha_{13}$) & -- & -- & 1.164 & -- & -- & -- & 1.039 \\
\hline
{\sc pow + rel} ($\alpha_{22}$) & 1.030 & 0.933 & 1.302 & 1.094 & 1.071 & 1.056 & 1.337 \\
{\sc pow + rel + xill} ($\alpha_{22}$) & -- & -- & 1.164 & -- & -- & -- & 1.040
\end{tabular}
\label{tabchi2}
\end{table*}

\begin{table*}
 \centering
 \caption{Best-fit values of the parameters in {\sc relxill\_nk} for $\alpha_{13}$ free and $\alpha_{22}=0$. $^*$ indicates that the parameter is frozen. $R_{\rm ref}$ is the reflection fraction and is calculated as the ratio between the flux of the relativistic reflection component and the sum of the fluxes of the power law and the relativistic reflection components in the energy range 0.6-10~keV.}
\vspace{0.2cm} 
\begin{tabular}{cccccccc}
Source & Ton~S180 & RBS~1124 & Ark~120 & Swift~J0501.9--3239 & 1H0419--577 & PKS 0558--504 & Fairall~9 \\
\hline
$q_{\rm in}$ & $> 9.7$ & $8_{-3}$ & $9.5^{+0.5}_{-1.1}$ & $> 9.86$ & $7.4^{+1.0}_{-1.4}$ & $> 9.4$ & $7.10^{+0.11}_{-0.30}$ \\
$q_{\rm out}$ & $3.0^*$ & $3.0^*$ & $3.5^{+0.3}_{-0.4}$ & $=q_{\rm in}$ & $=q_{\rm in}$ & $3.0^*$ & $3.06^{+0.05}_{-0.14}$ \\
$R_{\rm br}$~[$M$] & $3.15^{+0.18}_{-0.51}$ & $1.9^{+0.5}_{-0.3}$ & $3.6^{+0.3}_{-0.5}$ & -- & -- & $2.83^{+0.13}_{-0.14}$ & $3.04^{+0.06}_{-0.33}$ \\
$a_*$ & $0.996_{-0.005}$ & $0.989^{+0.004}_{-0.125}$ & $> 0.996$ & $0.9925^{+0.0022}_{-0.0017}$ & $> 0.992$ & $> 0.993$ & $0.9939_{-0.0008}$ \\
$\alpha_{13}$ & $0.01^{+0.02}_{-0.32}$ & $-0.7^{+0.1}_{-0.1}$ & $0.00^{+0.01}_{-0.08}$ & $0.00^{+0.03}_{-0.07}$ & $0.00^{+0.04}_{-0.14}$ & $0.03^{+0.02}_{-0.20}$ & $-0.7^{+0.3}_{-0.1}$ \\
$i$~[deg] & $37.4^{+2.0}_{-3.2}$ & $47^{+6}_{-13}$ & $25^{+4}_{-4}$ & $< 15$ & $71^{+3}_{-4}$ & $44^{+4}_{-3}$ & $35.1^{+0.8}_{-3.5}$ \\
$\log\xi$ & $3.23^{+0.04}_{-0.16}$ & $1.30^{+0.23}_{-0.35}$ & $2.97^{+0.04}_{-0.13}$ & $2.8^{+0.3}_{-0.3}$ & $0.69^{+0.17}_{-0.08}$ & $3.000^{+0.025}_{-0.120}$ & $2.419^{+0.292}_{-0.024}$ \\
$A_{\rm Fe}$ & $2.7^{+0.6}_{-0.5}$ & $1.8^{+0.7}_{-0.7}$ & $1.9^{+0.6}_{-0.7}$ & $1.7^{+1.6}_{-0.3}$ & $2.1^{+0.5}_{-0.6}$ & $4.7^{+2.6}_{-0.8}$ & $2.6^{+0.9}_{-0.7}$ \\
$\Gamma$ & $2.43^{+0.03}_{-0.03}$ & $1.97^{+0.02}_{-0.03}$ & $2.38^{+0.04}_{-0.03}$ & $2.311^{+0.061}_{-0.009}$ & $2.16^{+0.03}_{-0.04}$ & $2.321^{+0.013}_{-0.012}$ & $2.049^{+0.026}_{-0.004}$ \\
$E_{\rm cut}$ & $300^*$ & $300^*$ & $300^*$ & $300^*$ & $300^*$ & $300^*$ & $300^*$ \\
\hline
$R_{\rm ref}$ & $> 0.99$ & $0.24^{+0.21}_{-0.13}$ & $0.77^{+0.20}_{-0.16}$ & $0.68^{+0.21}_{-0.03}$ & $0.13^{+0.04}_{-0.03}$ & $0.29^{+0.26}_{-0.21}$ & $0.21^{+0.07}_{-0.04}$ \\
\hline
$\chi^2/\nu$ & 1352.33/1313 & 668.39/718 & 1404.44/1305 & 1352.65/1313 & 2489.64/2344 & 1380.52/1311 & 1299.57/1256 \\
& =1.0300 & =0.9309 & =1.0762 & =1.0302 & =1.0621 & =1.0530 & =1.0347 
\end{tabular}
\label{tabal13}
\end{table*}

\begin{table*}
 \centering
 \caption{Best-fit values of the parameters in {\sc relxill\_nk} for $\alpha_{13}=0$ and $\alpha_{22}$ free. $^*$ indicates that the parameter is frozen. $R_{\rm ref}$ is the reflection fraction and is calculated as the ratio between the flux of the relativistic reflection component and the sum of the fluxes of the power law and the relativistic reflection components in the energy range 0.6-10~keV.}
\vspace{0.2cm} 
\begin{tabular}{cccccccc}
Source & Ton~S180 & RBS~1124 & Ark~120 & Swift~J0501.9--3239 & 1H0419--577 & PKS 0558--504 & Fairall~9 \\
\hline
$q_{\rm in}$ & $> 9.4$ & $9_{-4}$ & $9.6_{-0.6}$ & $> 9.84$ & $7.4^{+1.5}_{-1.2}$ & $> 9.4$ & $> 9.4$ \\
$q_{\rm out}$ & $3^*$ & $3^*$ & $3.5^{+0.3}_{-0.3}$ & $=q_{\rm in}$ & $=q_{\rm in}$ & $3^*$ & $3.30^{+0.19}_{-0.31}$ \\
$R_{\rm br}$~[$M$] & $3.15^{+0.17}_{-0.10}$ & $1.72^{+1.37}_{-0.08}$ & $3.57^{+0.19}_{-0.32}$ & -- & -- & $2.79^{+0.17}_{-0.16}$ & $2.20^{+0.21}_{-0.16}$ \\
$a_*$ & $> 0.989$ & $0.997_{-0.004}$ & $> 0.995$ & $0.988^{+0.004}_{-0.004}$ & $> 0.994$ & $> 0.992$ & $0.992_{-0.011}$ \\
$\alpha_{22}$ & $-0.02^{+0.30}_{-0.04}$ & $1.2^{+0.8}_{-0.4}$ & $0.01^{+0.06}_{-0.03}$ & $0.11^{+0.05}_{-0.18}$ & $0.00^{+0.13}_{-0.04}$ & $-0.03^{+0.19}_{-0.02}$ & $1.3^{+0.2}_{-0.4}$ \\
$i$~[deg] & $36.7^{+3.2}_{-1.6}$ & $45.0^{+1.7}_{-2.0}$ & $25^{+4}_{-3}$ & $< 15$ & $71^{+4}_{-4}$ & $44^{+3}_{-4}$ & $35.5^{+1.4}_{-2.1}$ \\
$\log\xi$ & $3.23^{+0.05}_{-0.09}$ & $1.4^{+0.4}_{-0.3}$ & $2.97^{+0.05}_{-0.21}$ & $2.8^{+0.3}_{-0.3}$ & $0.69^{+0.09}_{-0.14}$ & $2.997^{+0.015}_{-0.116}$ & $2.44^{+0.13}_{-0.09}$ \\
$A_{\rm Fe}$ & $3.0^{+1.8}_{-1.6}$ & $1.8^{+0.7}_{-0.8}$ & $1.9^{+0.9}_{-0.8}$ & $1.47^{+0.72}_{-0.07}$ & $2.1^{+0.5}_{-0.6}$ & $4.7^{+2.6}_{-0.8}$ & $2.8^{+0.7}_{-0.6}$ \\
$\Gamma$ & $2.43^{+0.06}_{-0.03}$ & $1.95^{+0.06}_{-0.03}$ & $2.37^{+0.04}_{-0.03}$ & $2.304^{+0.019}_{-0.056}$ & $2.16^{+0.03}_{-0.04}$ & $2.318^{+0.016}_{-0.008}$ & $2.049^{+0.021}_{-0.018}$ \\
$E_{\rm cut}$ & $300^*$ & $300^*$ & $300^*$ & $300^*$ & $300^*$ & $300^*$ & $300^*$ \\
\hline
$R_{\rm ref}$ & $> 0.99$ & $0.23^{+0.09}_{-0.13}$ & $0.76^{+0.15}_{-0.12}$ & $0.65^{+0.10}_{-0.06}$ & $0.18^{+0.04}_{-0.04}$ & $0.29^{+0.13}_{-0.22}$ & $0.25^{+0.04}_{-0.03}$ \\
\hline
$\chi^2/\nu$ & 1353.10/1313 & 669.63/718 & 1404.33/1305 & 1354.16/1313 & 2489.66/2344 & 1380.60/1311 & 1298.17/1256 \\
& =1.0305 & =0.9326 & =1.0761 & =1.0313 & =1.0621 & =1.0531 & =1.0336 
\end{tabular}
\label{tabal22}
\end{table*}

\begin{figure*}[t]
\begin{center}
\includegraphics[width=8.5cm,trim={0.5cm 0 3cm 19cm},clip]{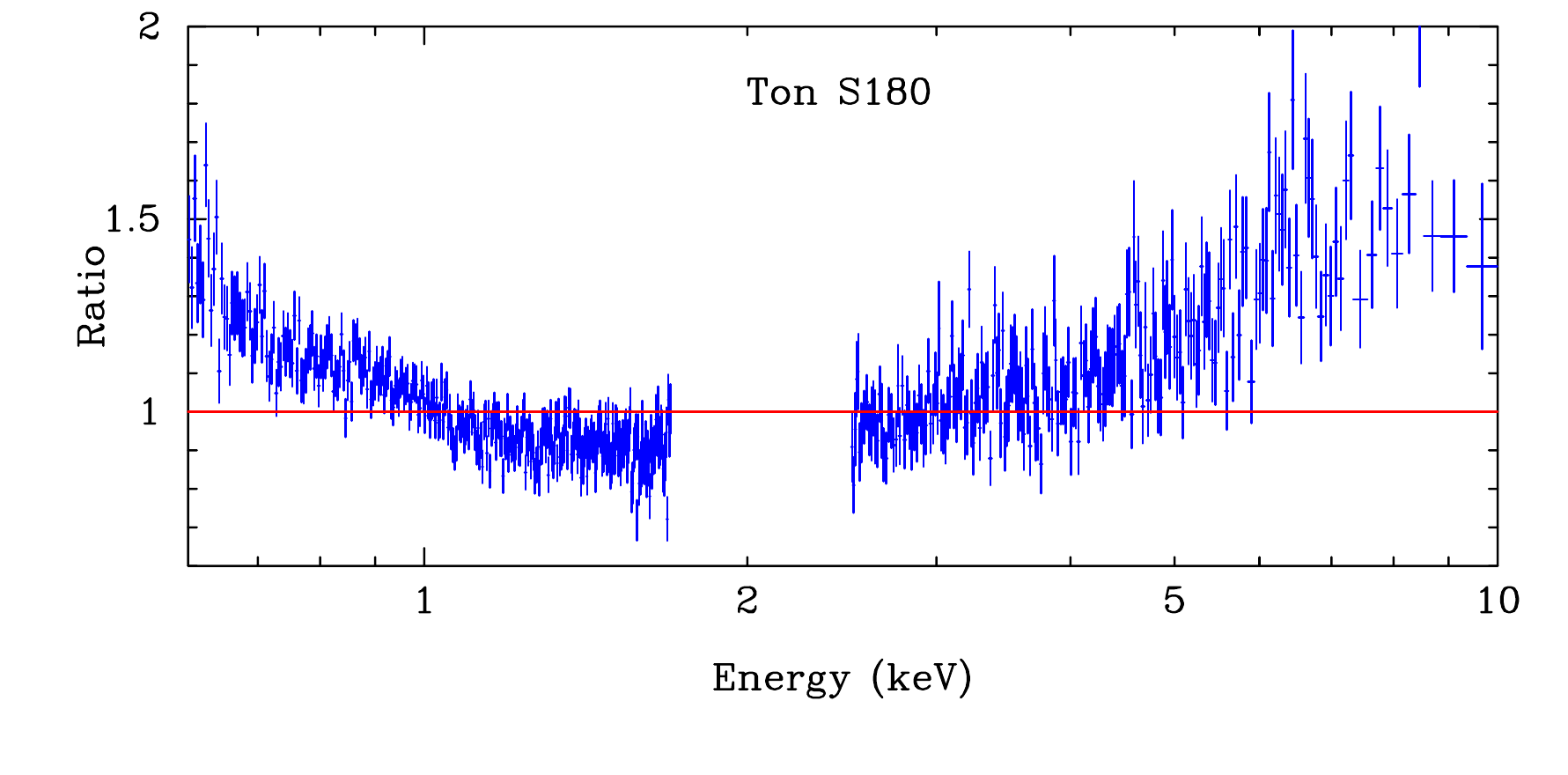}
\includegraphics[width=8.5cm,trim={0.5cm 0 3cm 19cm},clip]{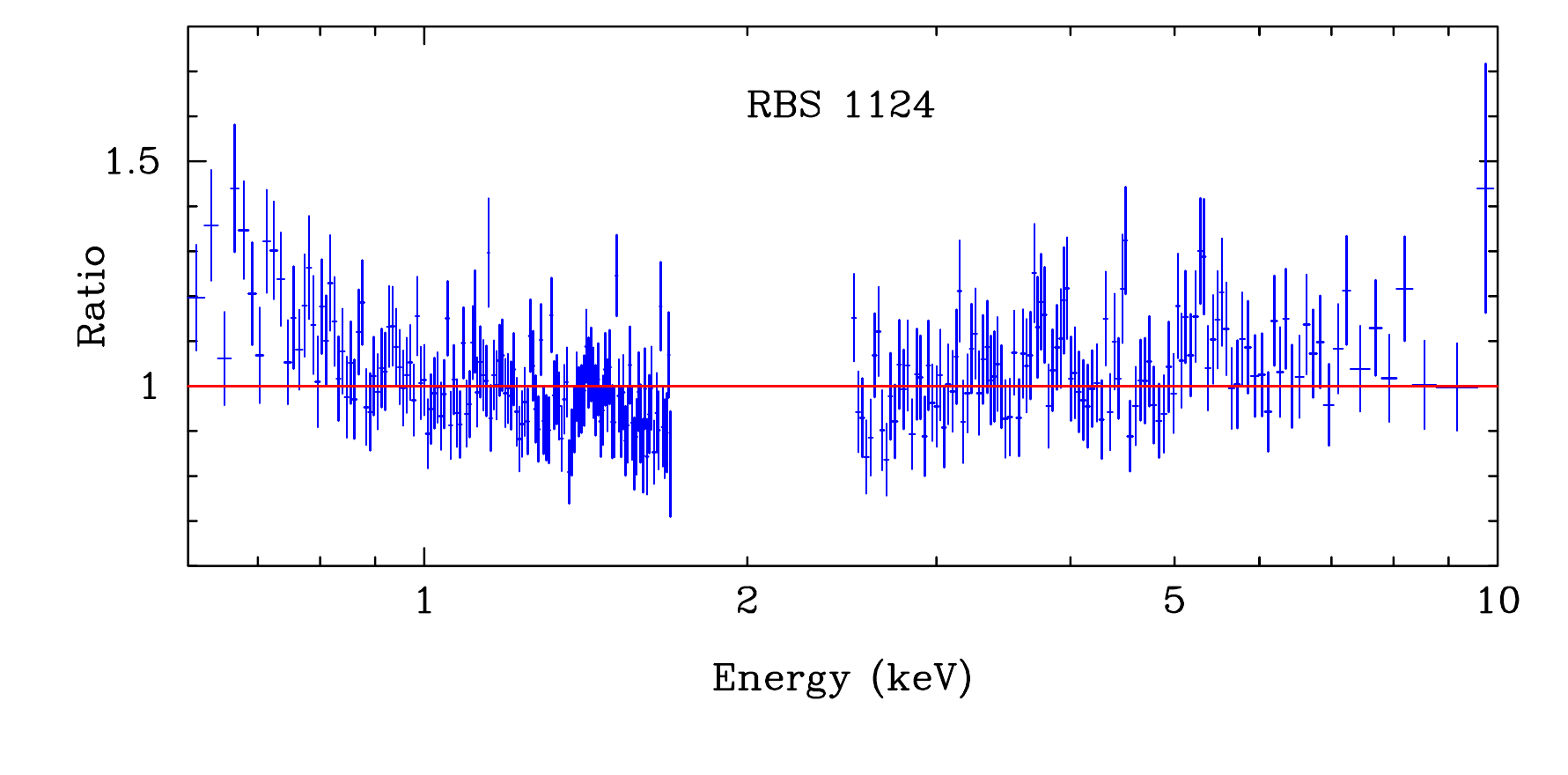} \\
\includegraphics[width=8.5cm,trim={0.5cm 0 3cm 19cm},clip]{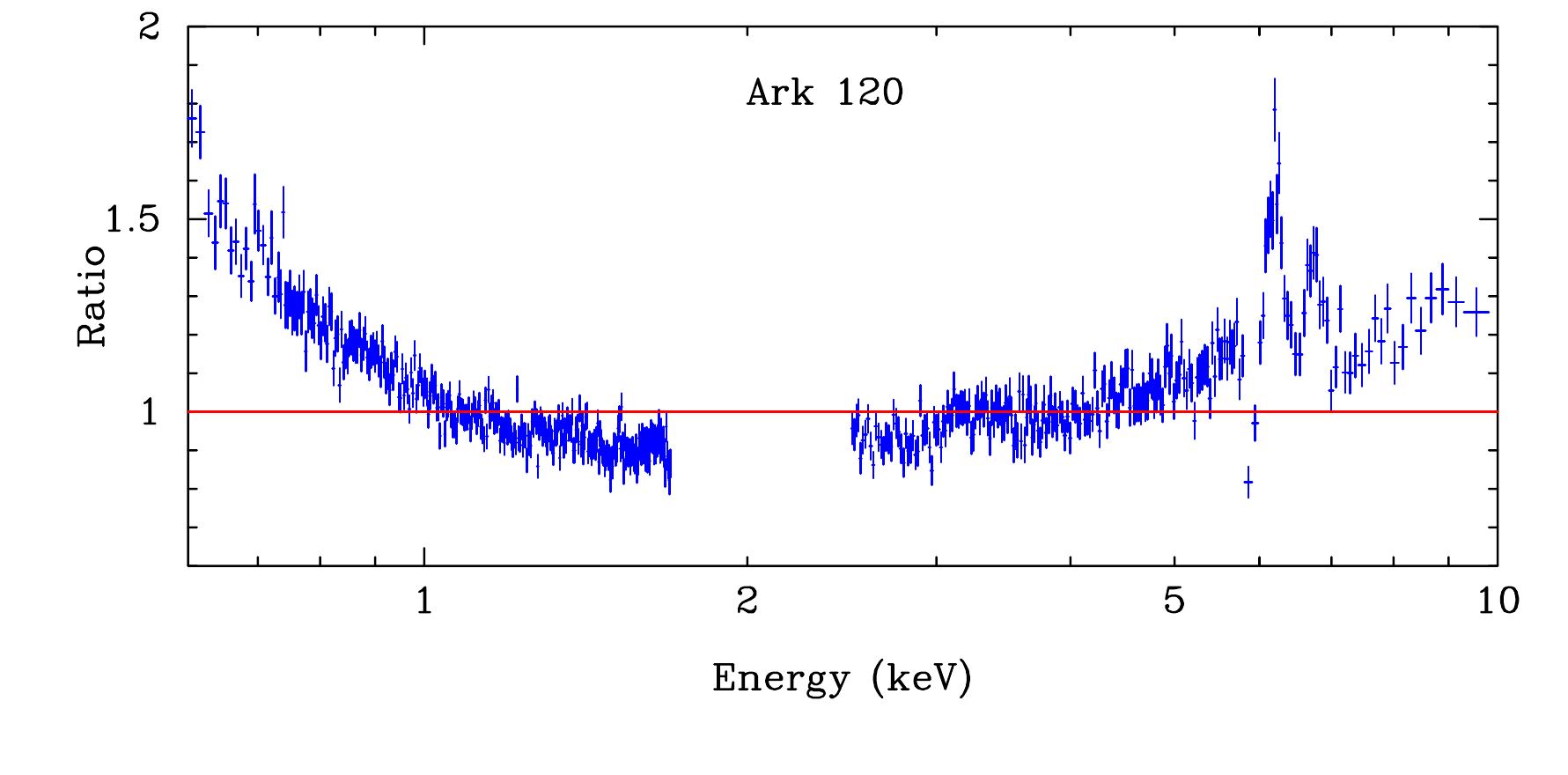}
\includegraphics[width=8.5cm,trim={0.5cm 0 3cm 19cm},clip]{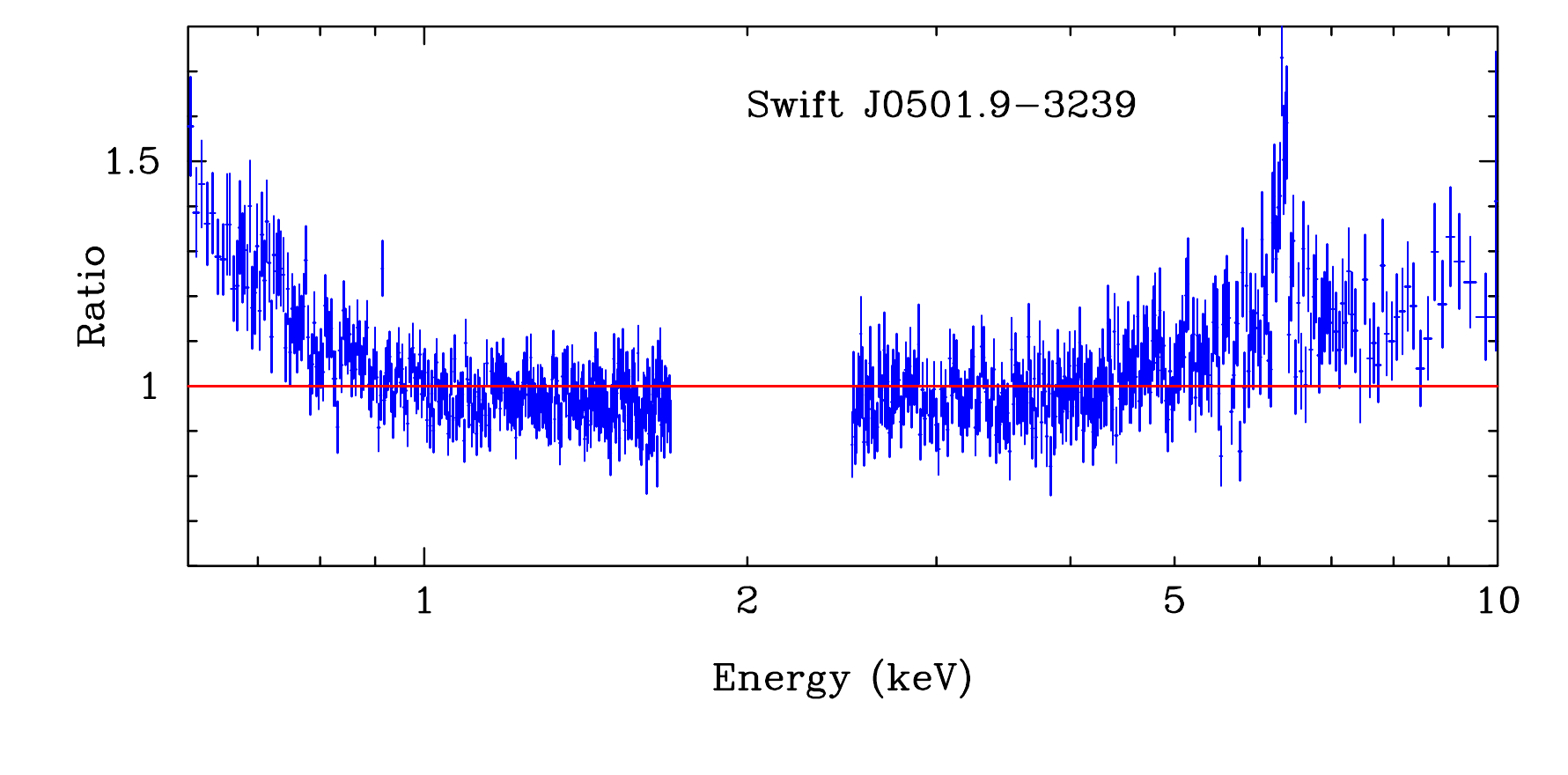} \\
\includegraphics[width=8.5cm,trim={0.5cm 0 3cm 19cm},clip]{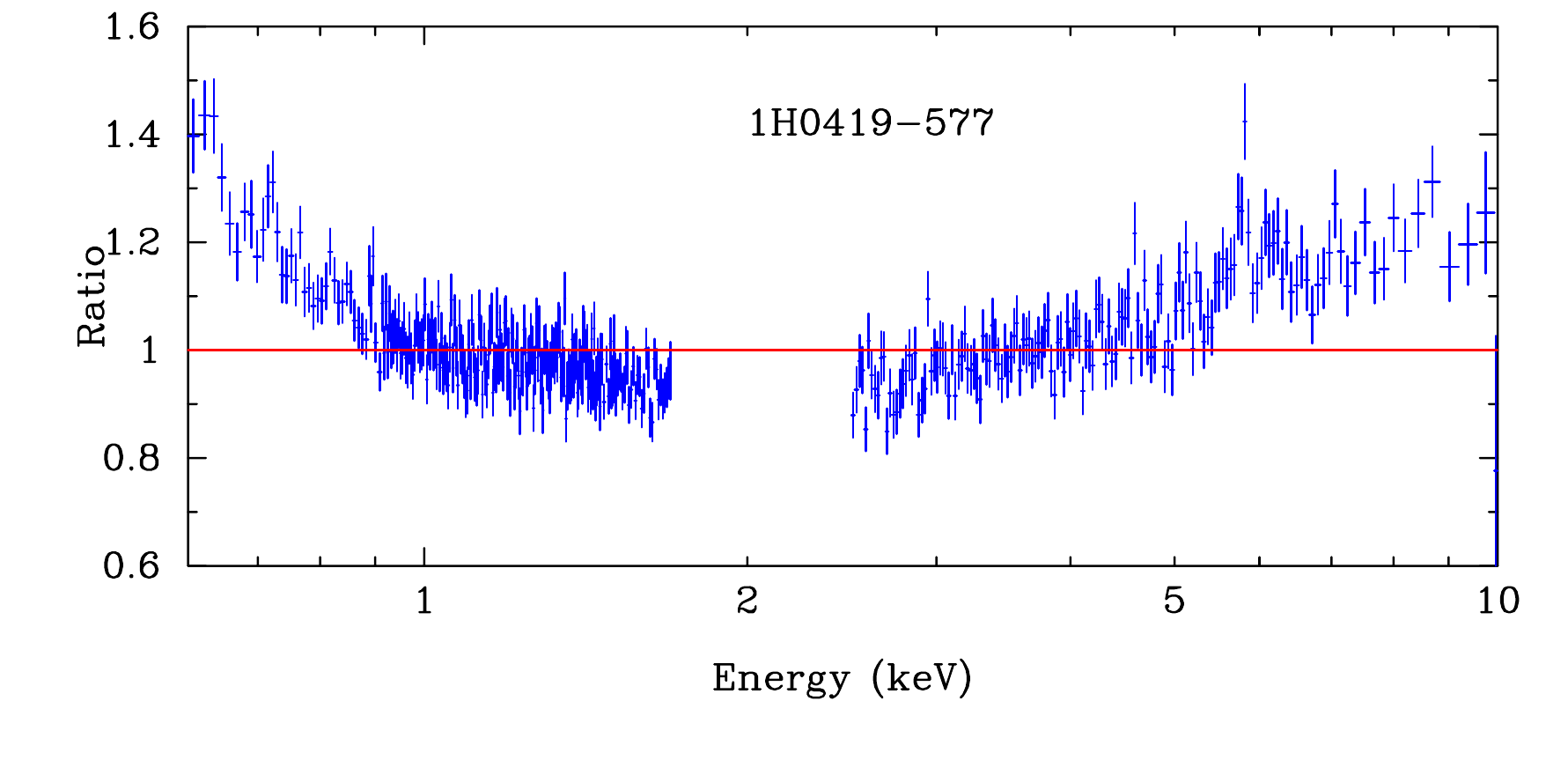}
\includegraphics[width=8.5cm,trim={0.5cm 0 3cm 19cm},clip]{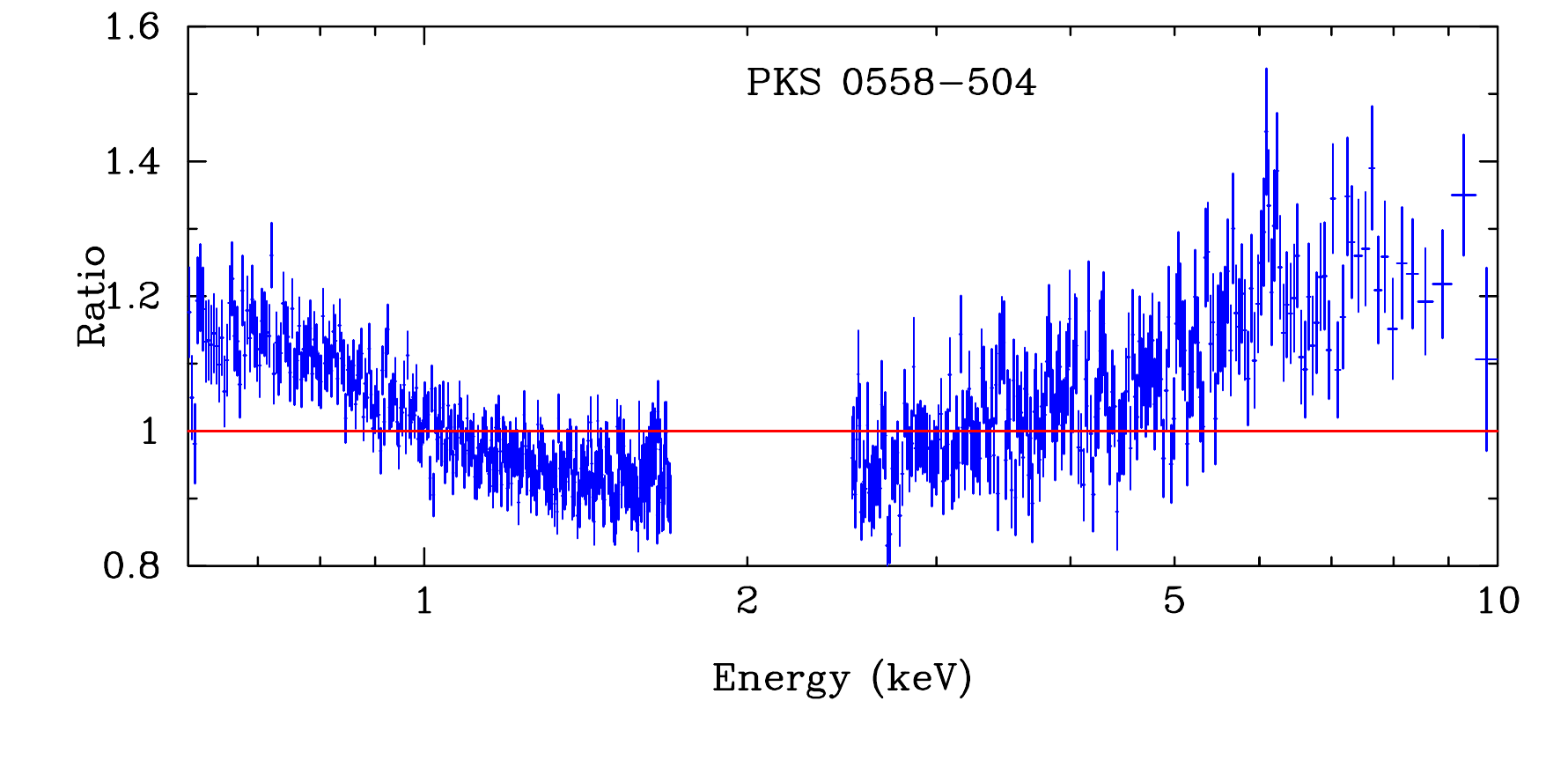} \\
\includegraphics[width=8.5cm,trim={0.5cm 0 3cm 19cm},clip]{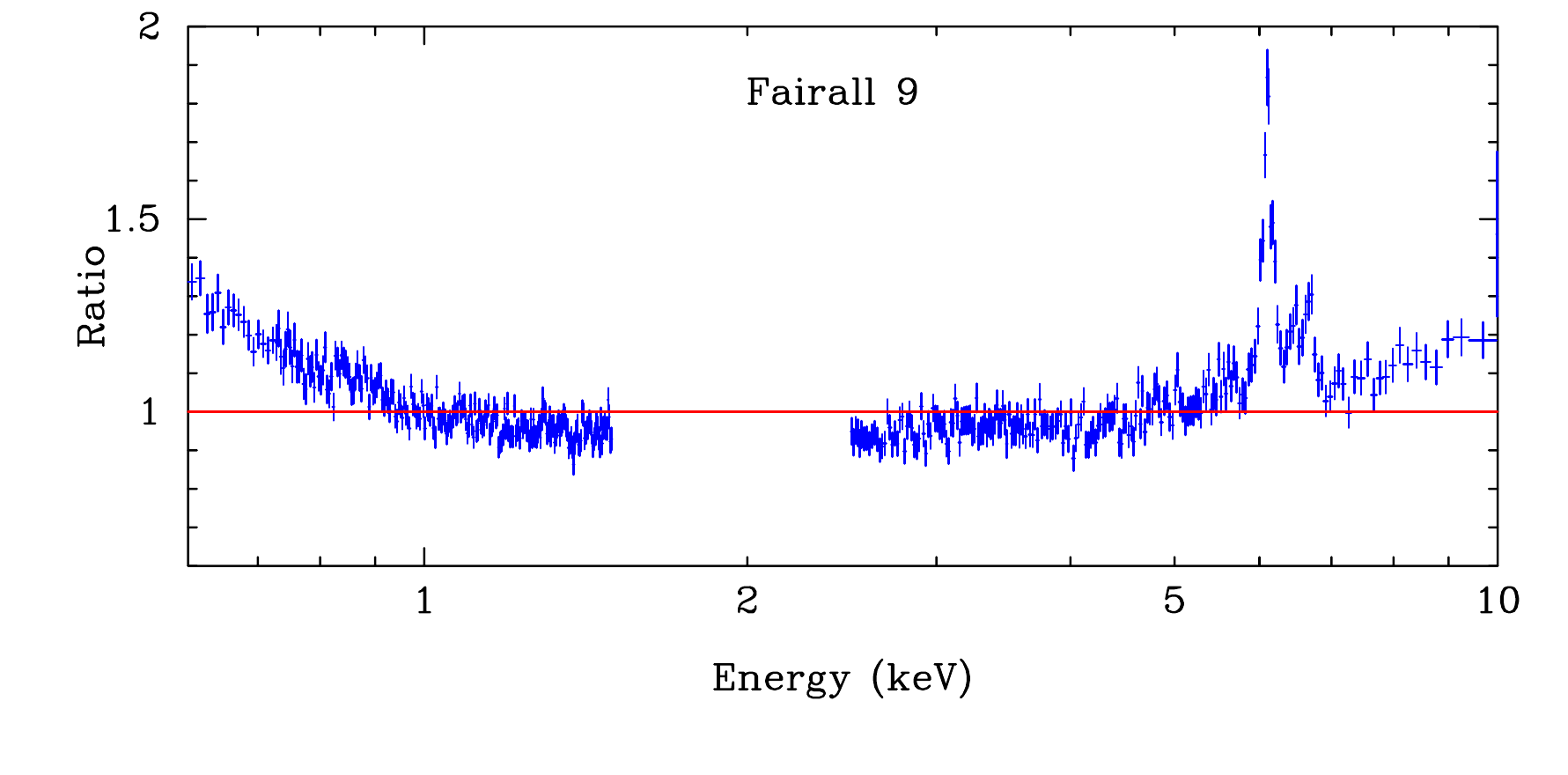}
\end{center}
\vspace{-0.7cm}
\caption{Data to best-fit model ratios for our seven sources when the spectra are described by a power law component only. \label{r-p}}
\end{figure*}

\begin{figure*}[t]
\begin{center}
\includegraphics[width=8.5cm,trim={0.5cm 0 3cm 18cm},clip]{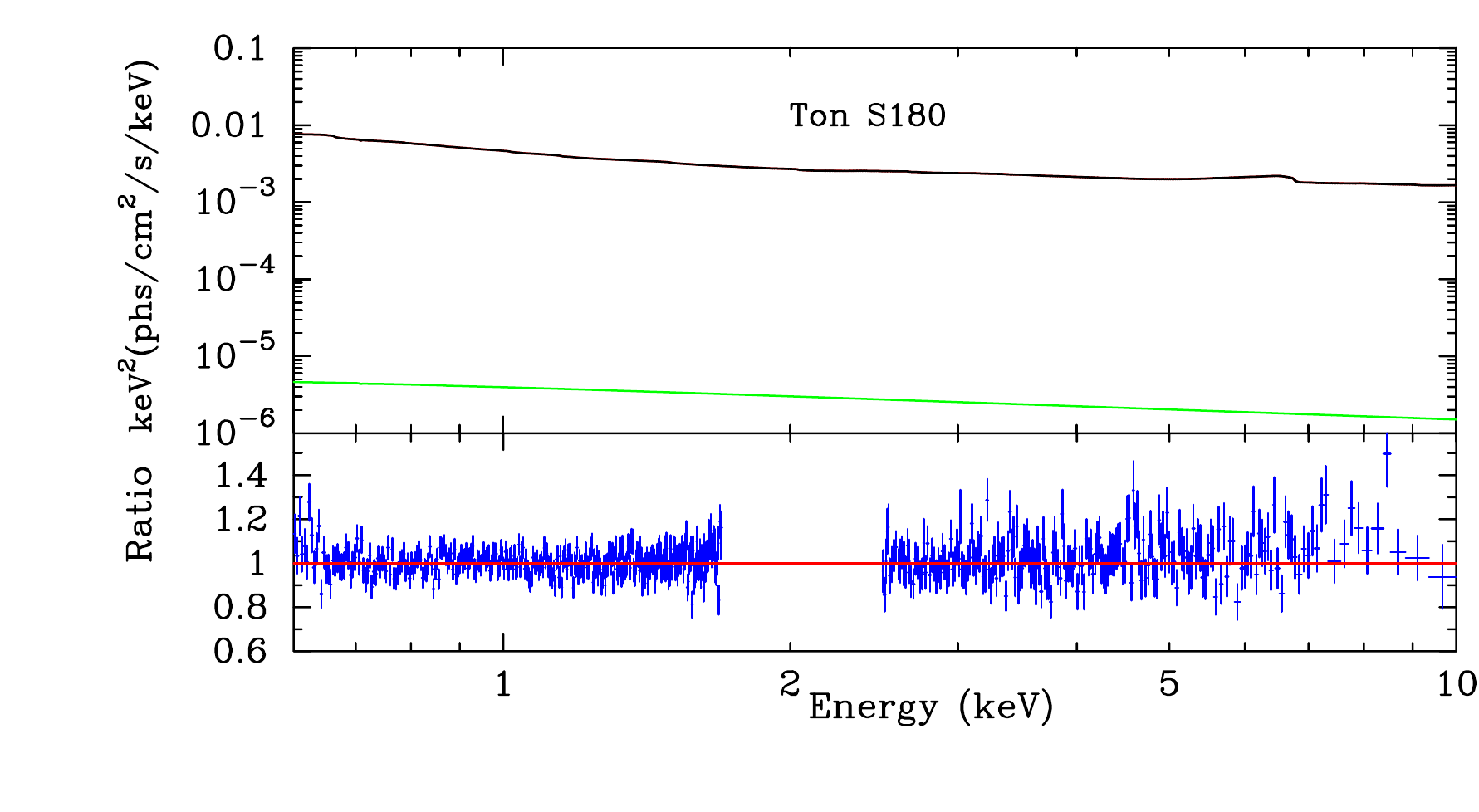}
\includegraphics[width=8.5cm,trim={0.5cm 0 3cm 18cm},clip]{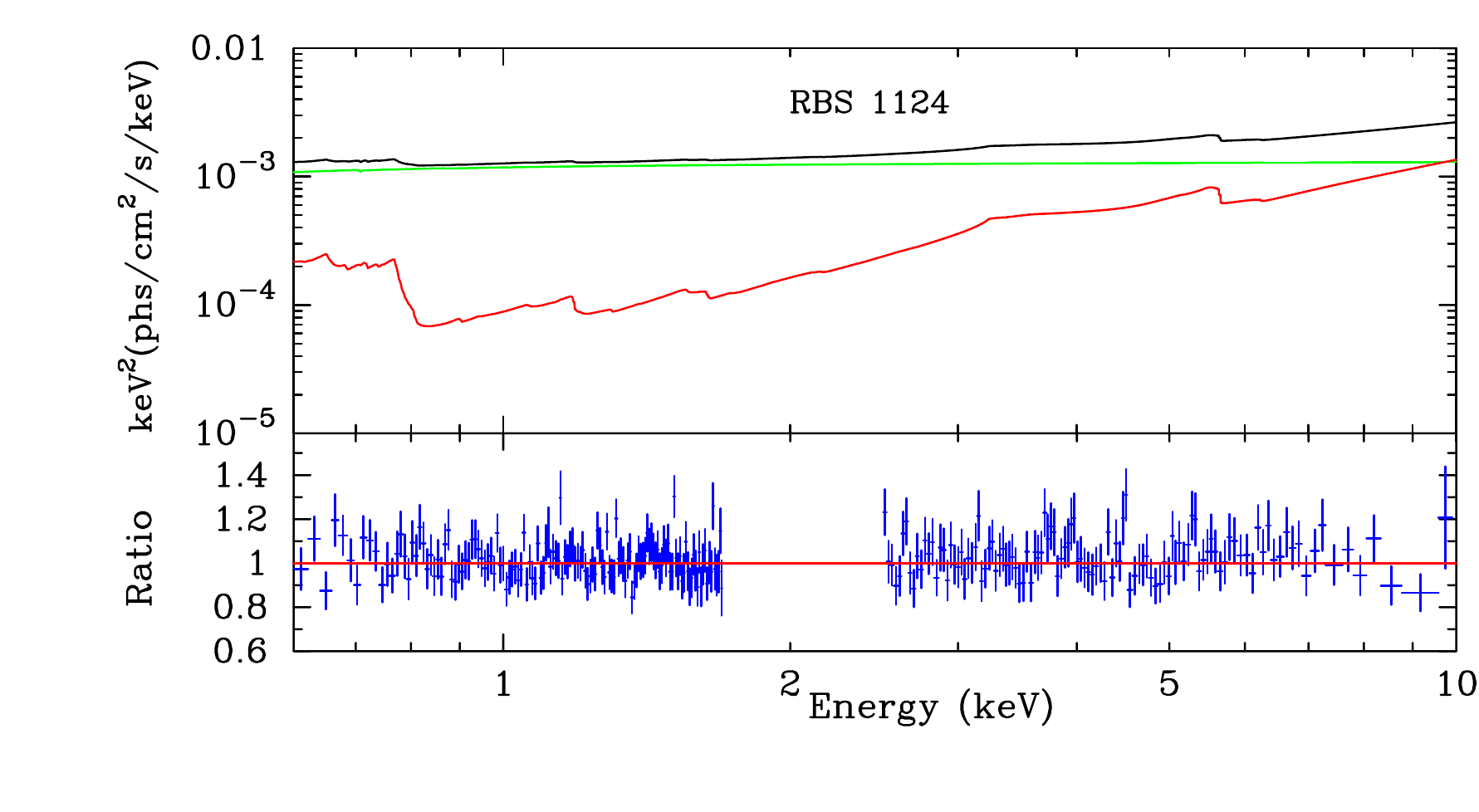} \\
\includegraphics[width=8.5cm,trim={0.5cm 0 3cm 18cm},clip]{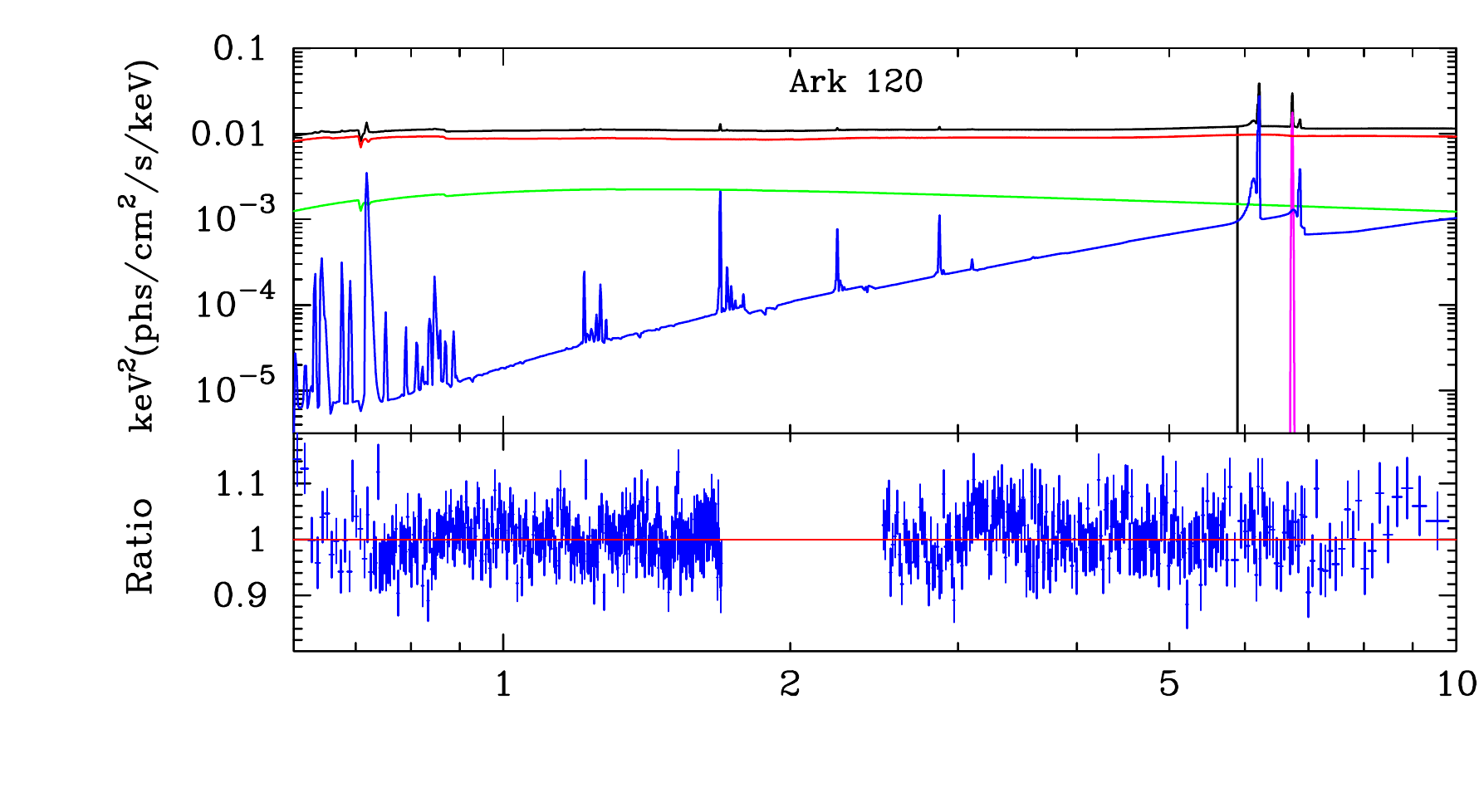}
\includegraphics[width=8.5cm,trim={0.5cm 0 3cm 18cm},clip]{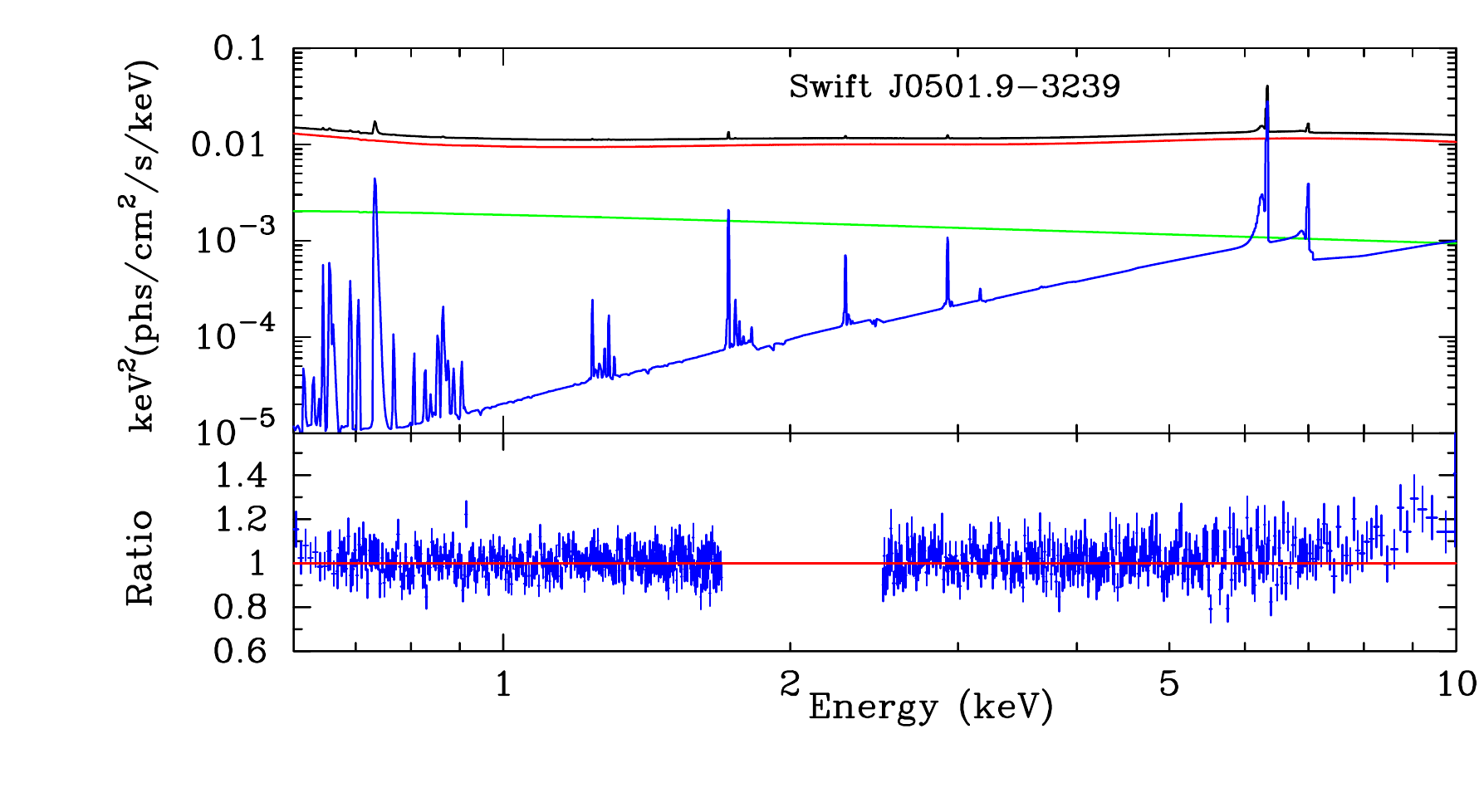} \\
\includegraphics[width=8.5cm,trim={0.5cm 0 3cm 18cm},clip]{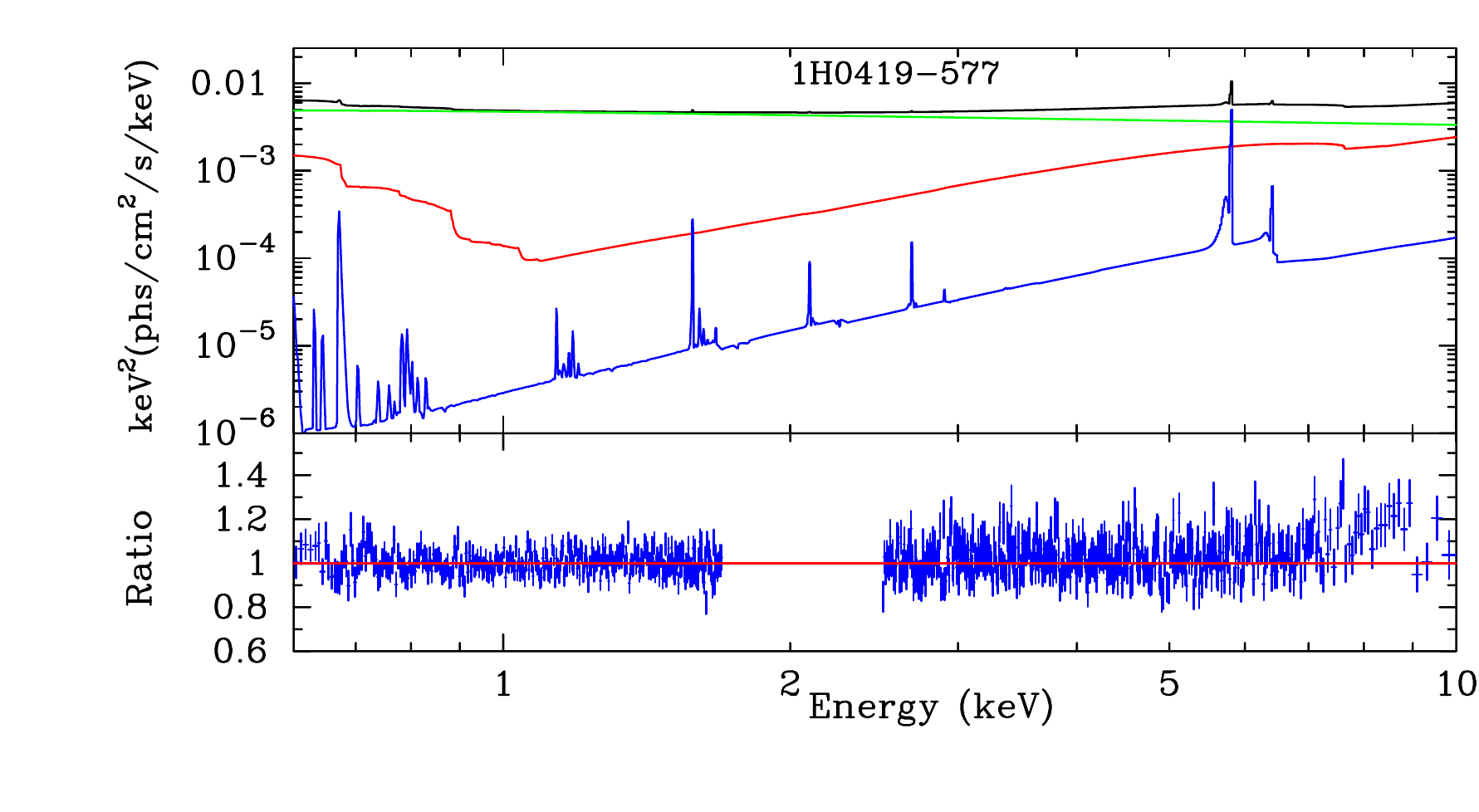}
\includegraphics[width=8.5cm,trim={0.5cm 0 3cm 18cm},clip]{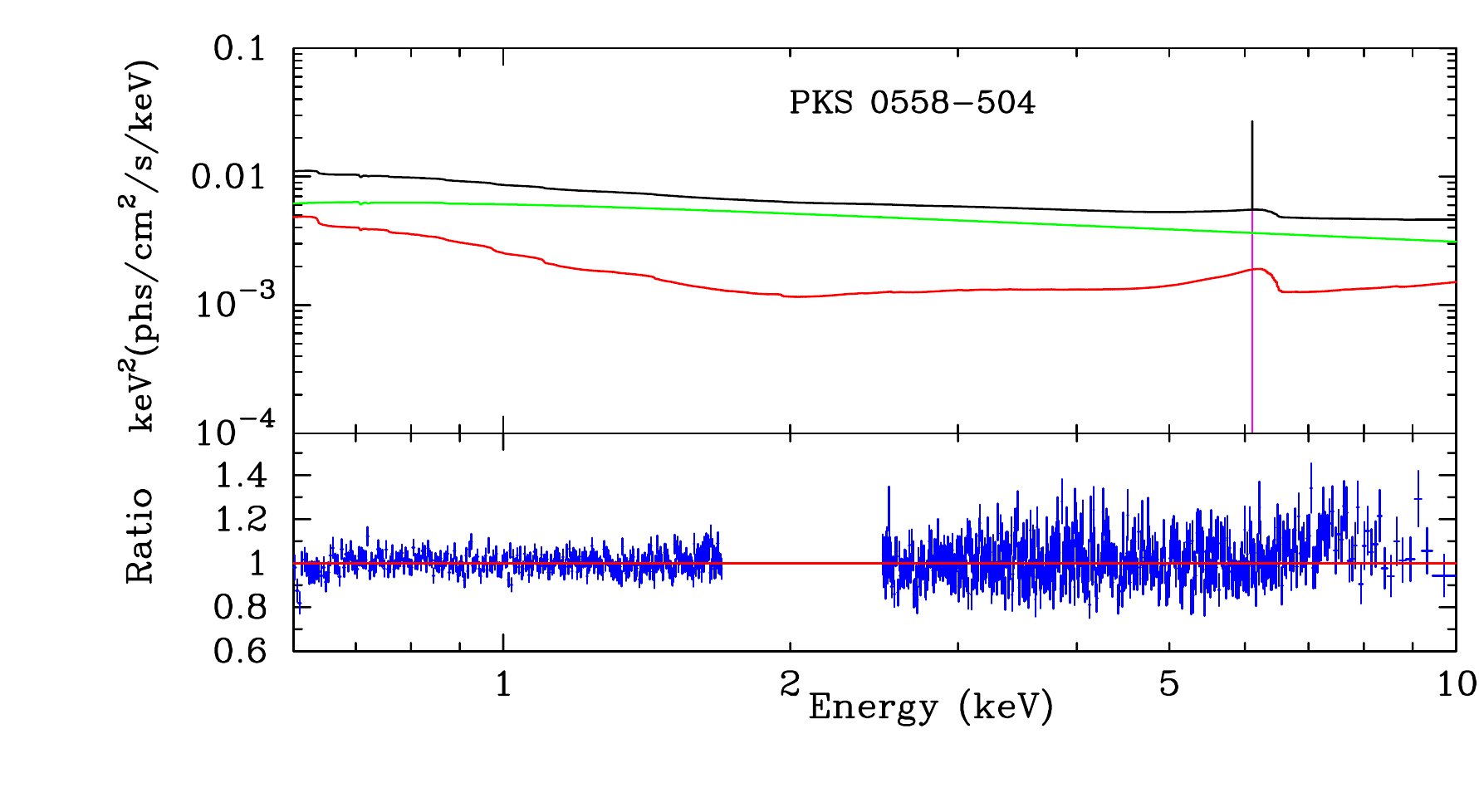} \\
\includegraphics[width=8.5cm,trim={0.5cm 0 3cm 18cm},clip]{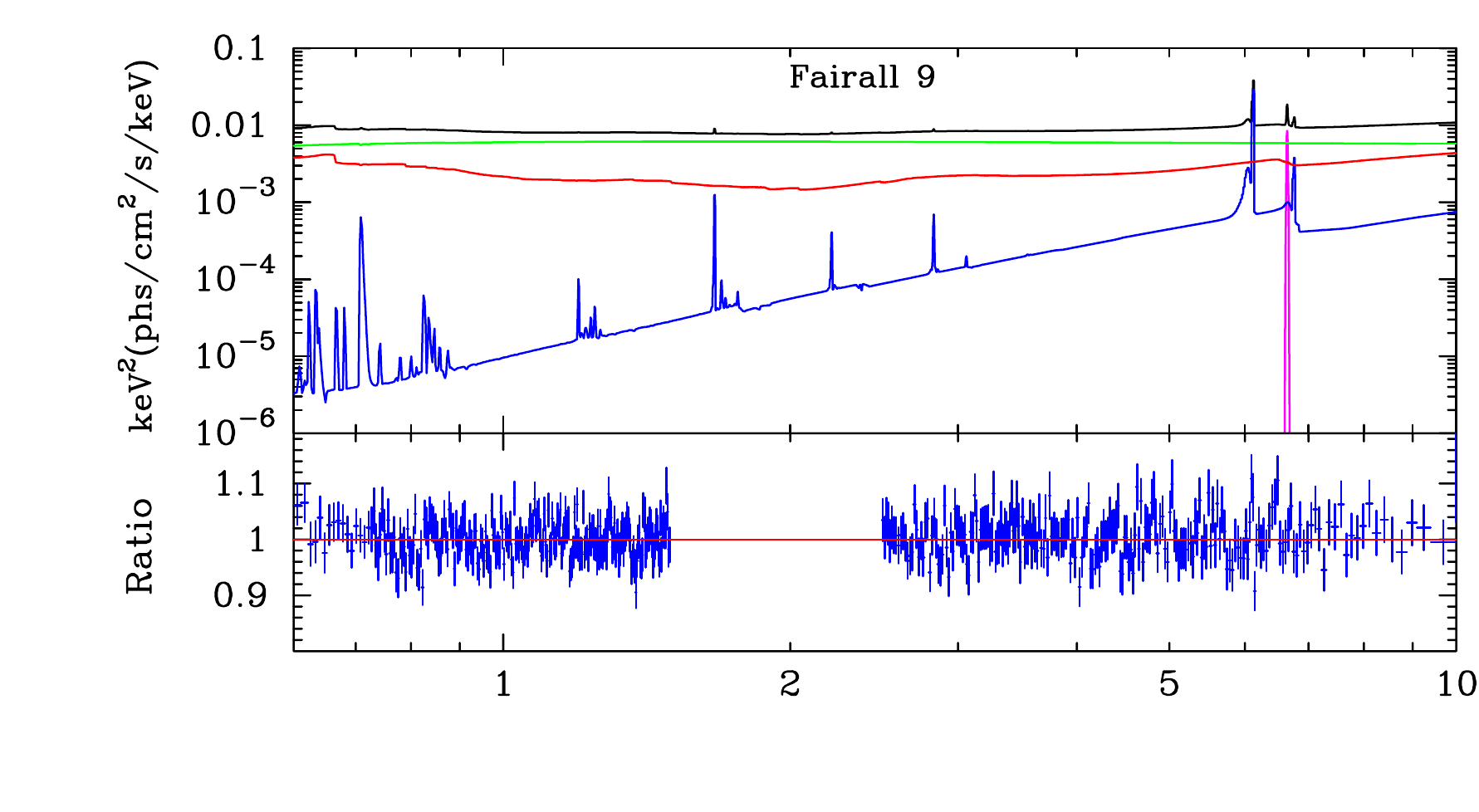}
\end{center}
\vspace{-0.7cm}
\caption{Spectra of the best fit models with the corresponding components (upper panels) and data to best-fit model ratios (lower panels) for our seven sources and the full models with $\alpha_{13}$ free and $\alpha_{22} = 0$. The total spectra are in black, power law components from the coronas are in green, relativistic reflection components from disks are in red, non-relativistic reflection components from distant reflectors are in blue, narrow lines are in magenta. Note that in Ton~S180 the total flux (black) and the relativistic reflection component (red) overlap. \label{r-a13}}
\end{figure*}

\begin{figure*}[t]
\begin{center}
\includegraphics[width=8.5cm,trim={0.5cm 0 3cm 18cm},clip]{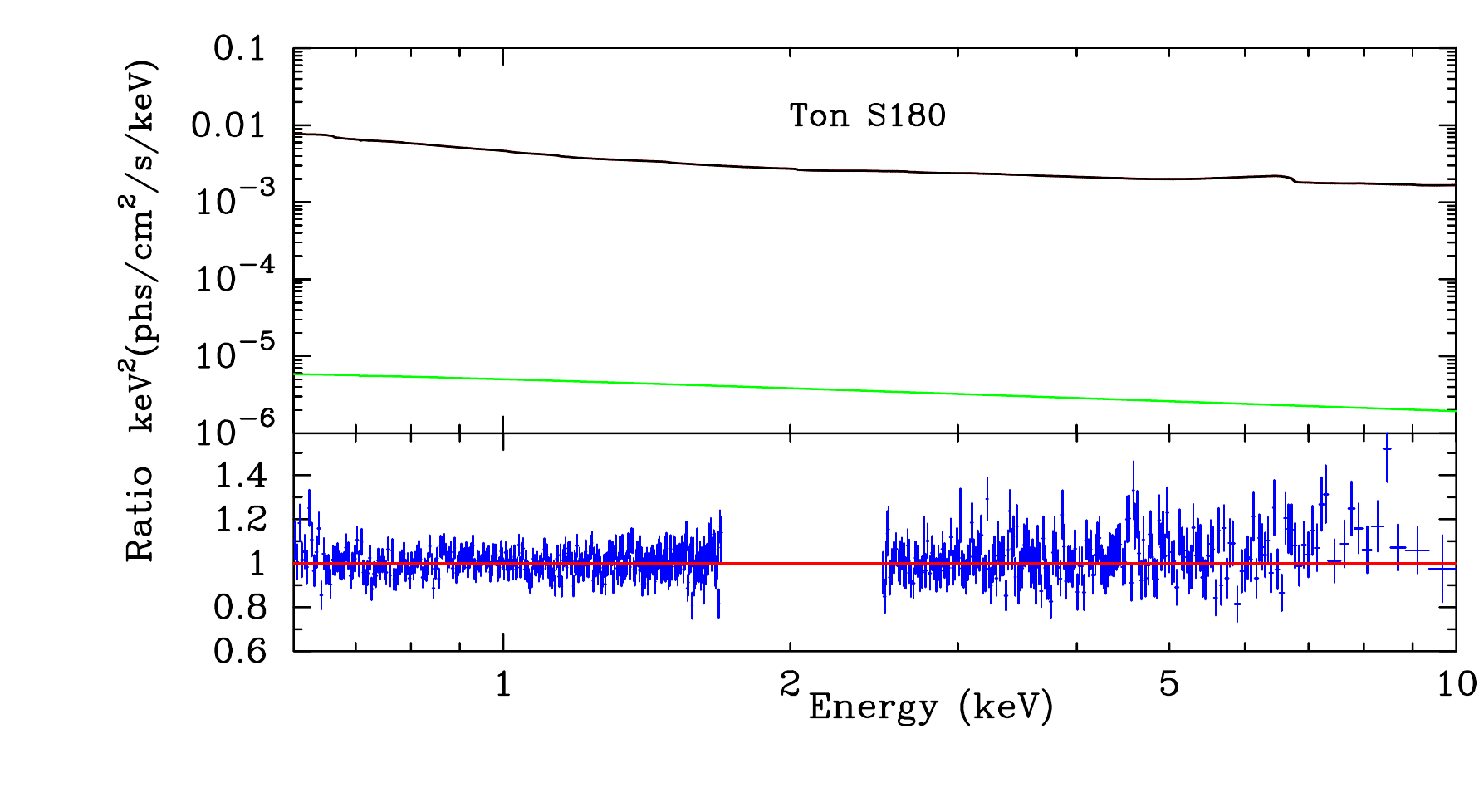}
\includegraphics[width=8.5cm,trim={0.5cm 0 3cm 18cm},clip]{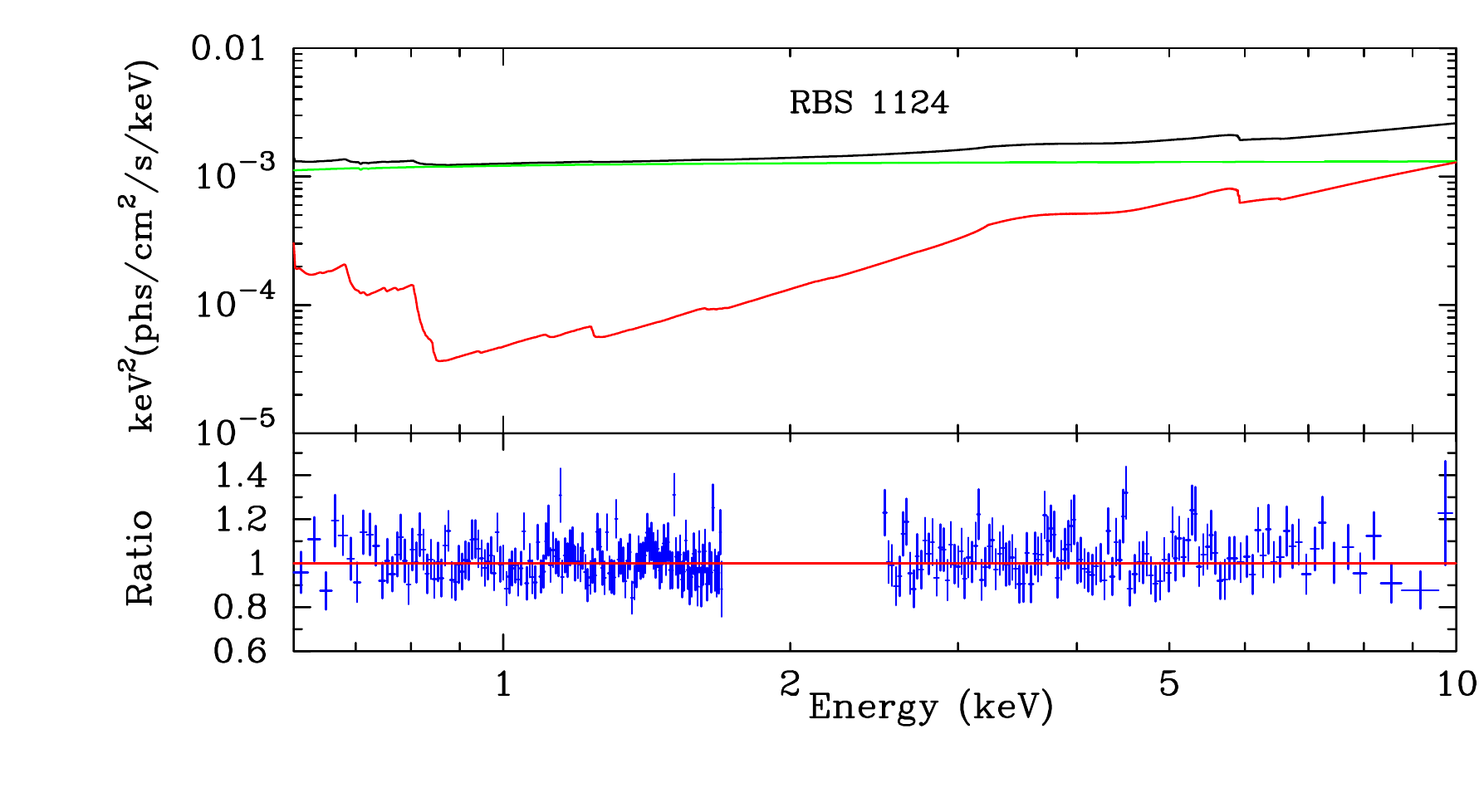} \\
\includegraphics[width=8.5cm,trim={0.5cm 0 3cm 18cm},clip]{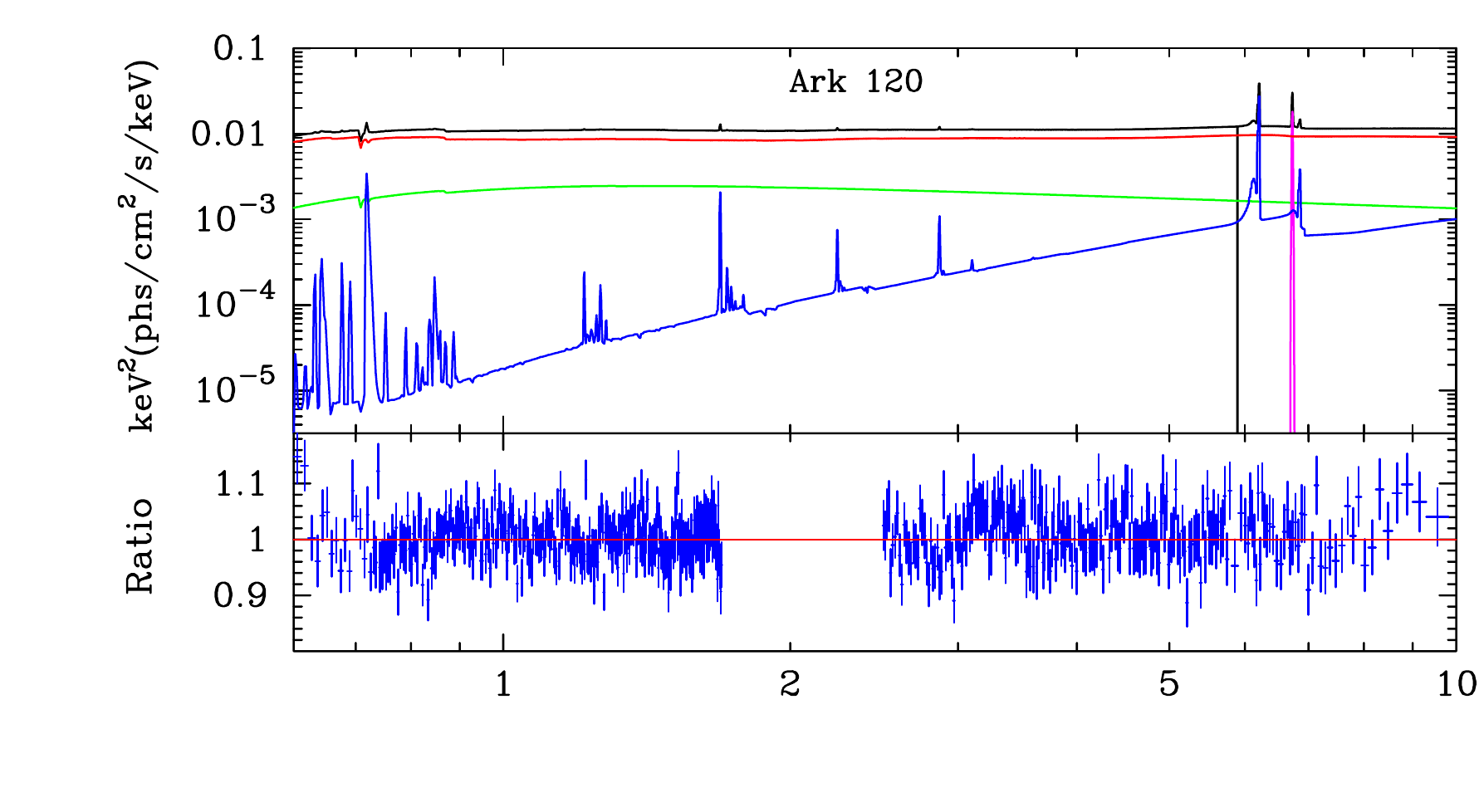}
\includegraphics[width=8.5cm,trim={0.5cm 0 3cm 18cm},clip]{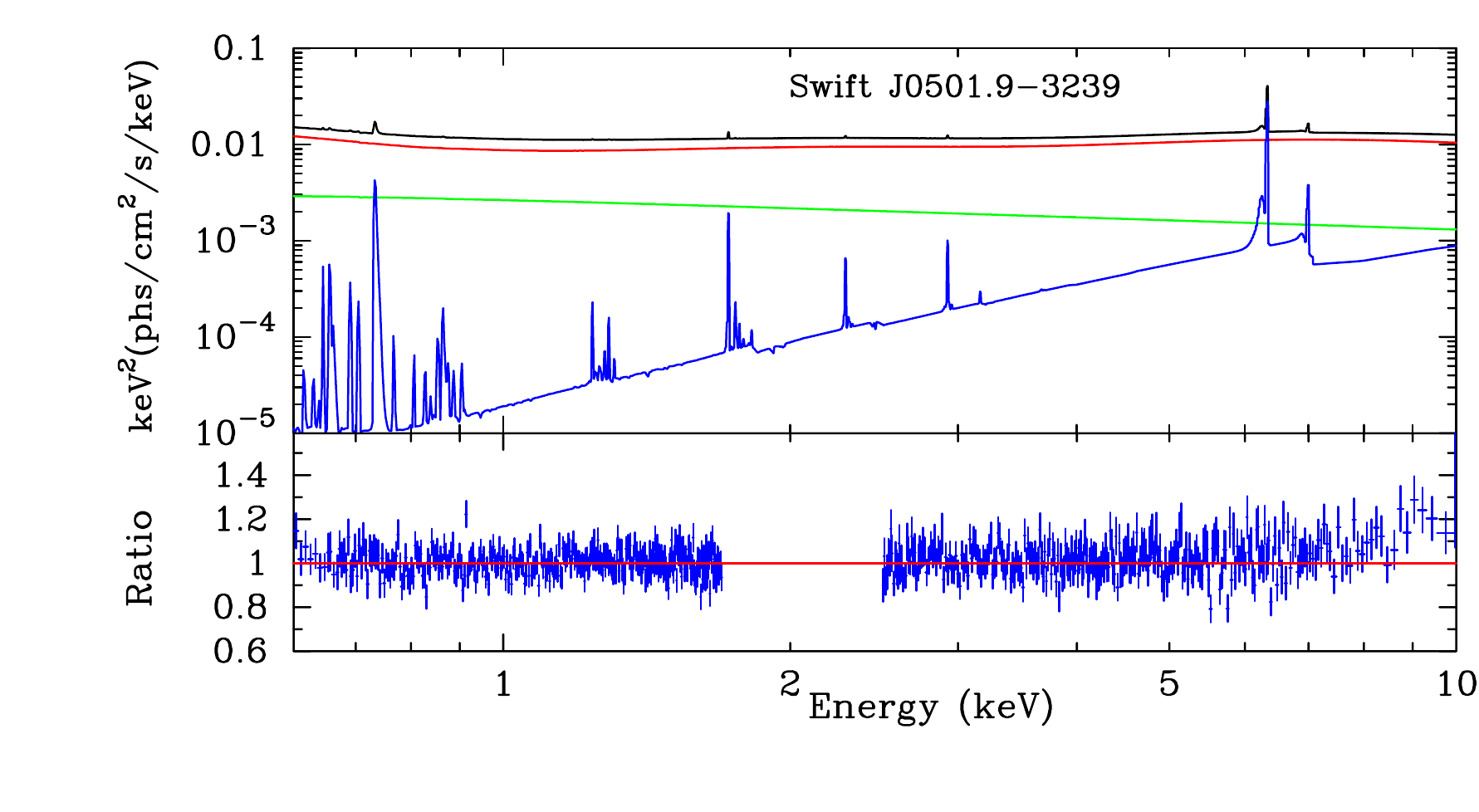} \\
\includegraphics[width=8.5cm,trim={0.5cm 0 3cm 18cm},clip]{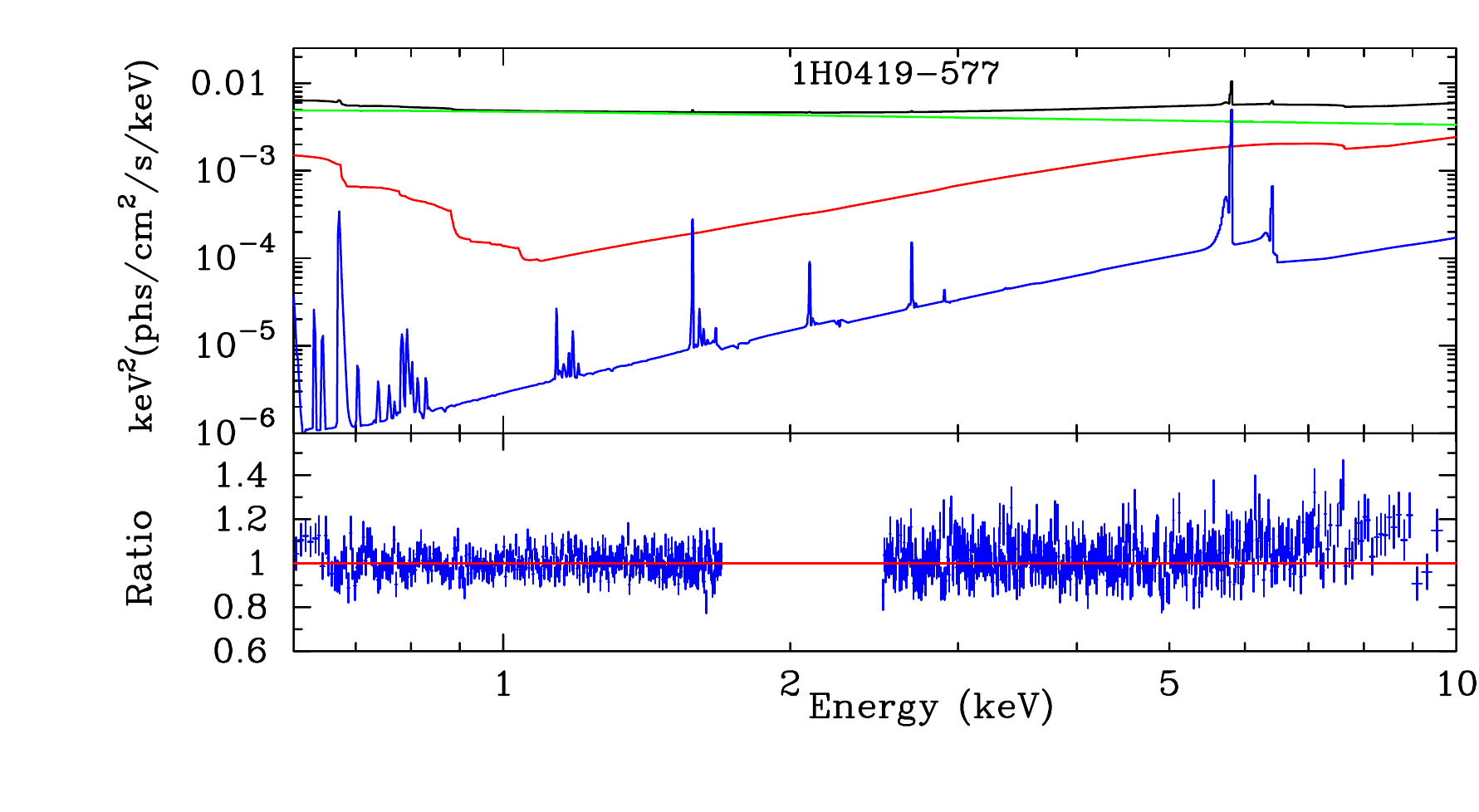}
\includegraphics[width=8.5cm,trim={0.5cm 0 3cm 18cm},clip]{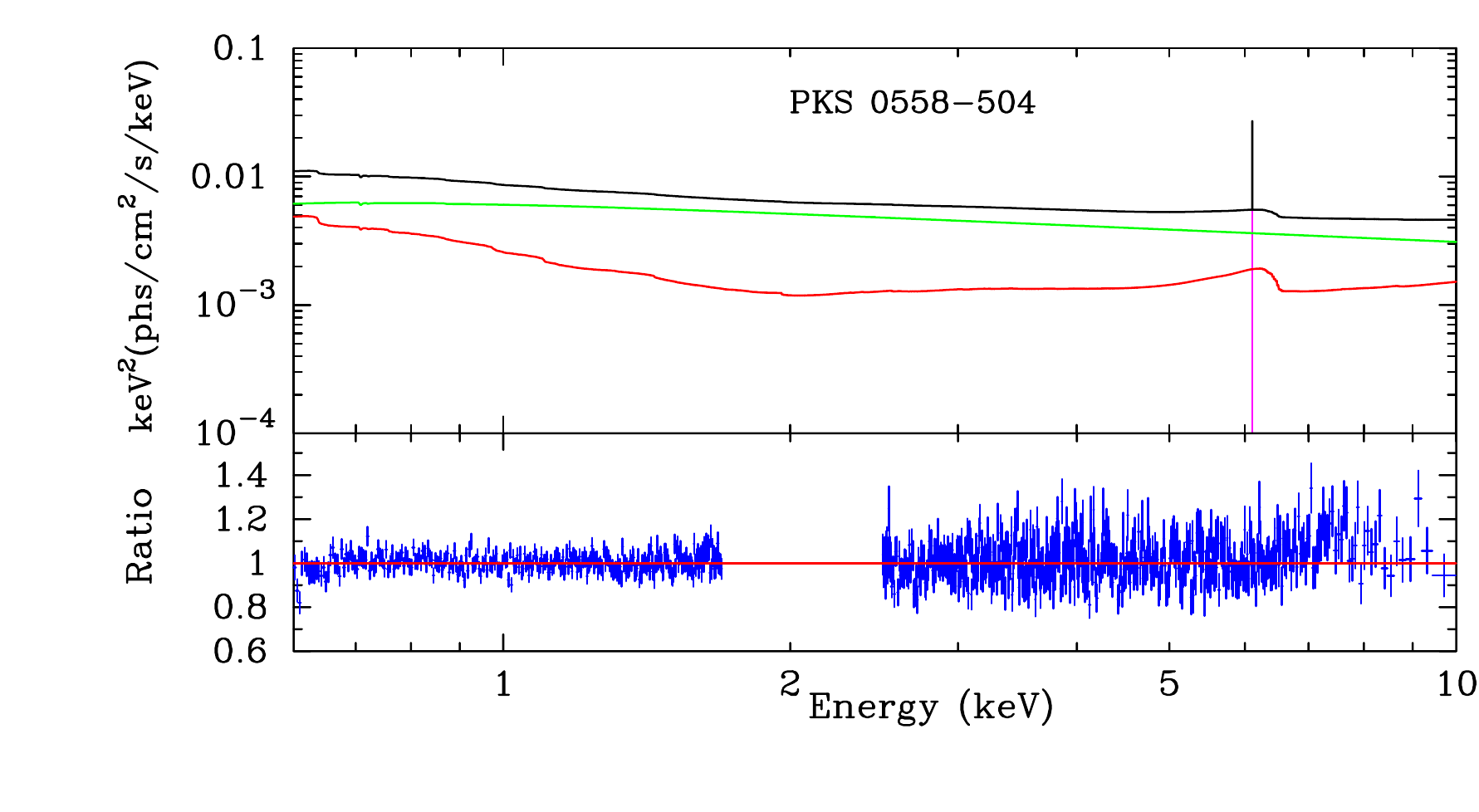} \\
\includegraphics[width=8.5cm,trim={0.5cm 0 3cm 18cm},clip]{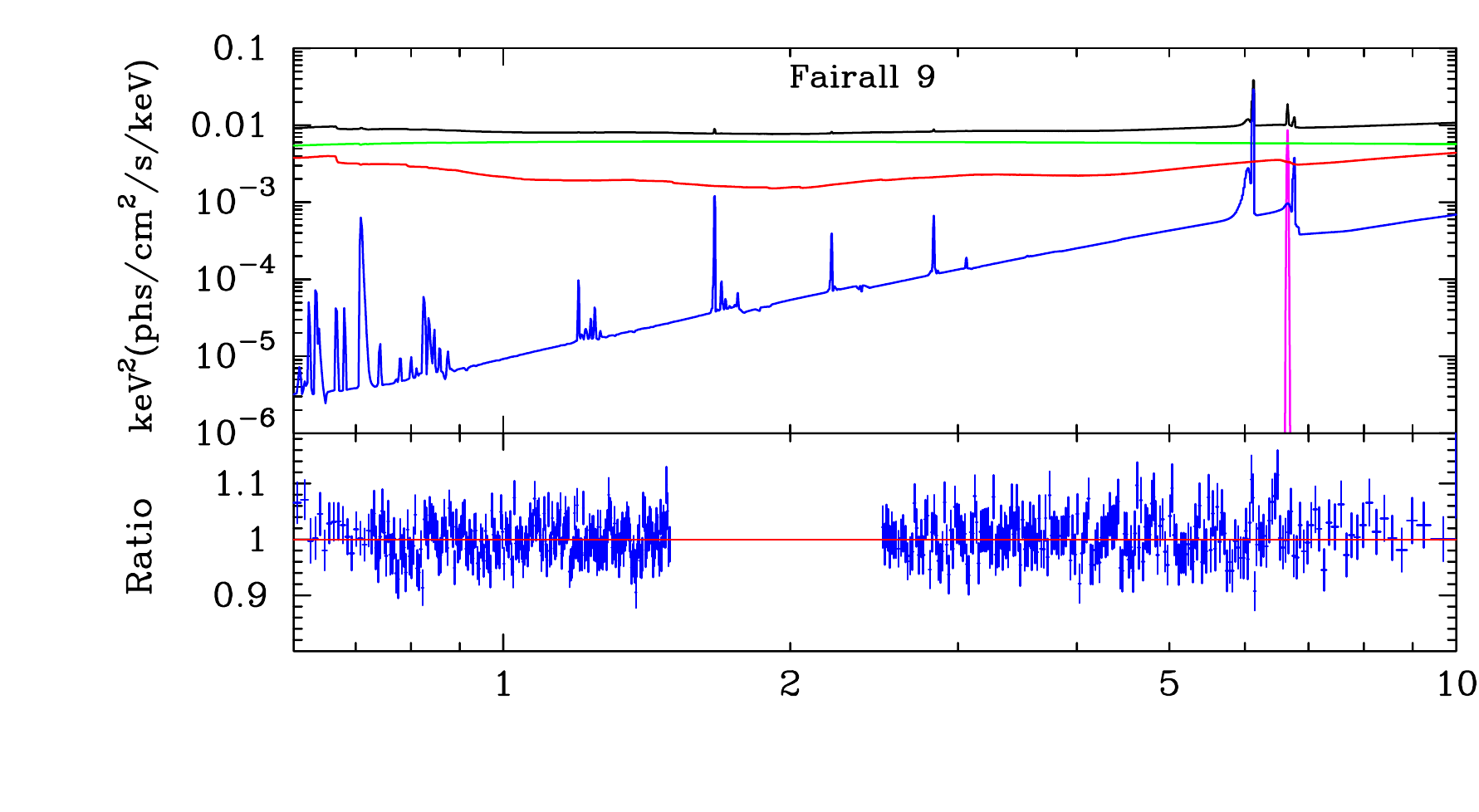}
\end{center}
\vspace{-0.7cm}
\caption{As in Fig.~\ref{r-a13} for the models with $\alpha_{13} = 0$ and $\alpha_{22}$ free. The total spectra are in black, power law components from the coronas are in green, relativistic reflection components from disks are in red, non-relativistic reflection components from distant reflectors are in blue, narrow lines are in magenta. Note that in Ton~S180 the total flux (black) and the relativistic reflection component (red) overlap. \label{r-a22}}
\end{figure*}

For every source, we start with the simplest model {\sc tbabs*zpowrlw}, which describes the spectrum of the corona taking the Galactic absorption into account. Fig.~\ref{r-p} shows the data to best-fit model ratios for the selected sources. All ratio plots present an excess of counts at low energies and, in most cases, a broad iron line at 5-7~keV. We thus add a relativistic reflection component: the new model is {\sc tbabs*(zpowrlw + relxill\_nk)} and improves the quality of the fit, as we can see from Tab.~\ref{tabchi2}. For some sources, we already get a good fit and we stop here. For other sources, we do not yet get a good fit and we add additional components, i.e. a non-relativistic reflection spectrum or some narrow emission lines. Fig.~\ref{r-a13} and Fig.~\ref{r-a22} show the spectra with the corresponding components (upper panels) and the data to best-fit model ratios (lower panels) for the final model of every source. In Fig.~\ref{r-a13}, $\alpha_{13}$ is free and $\alpha_{22} = 0$. In Fig.~\ref{r-a22}, we have the opposite case: $\alpha_{13} = 0$ and $\alpha_{22}$ is free. Tab.~\ref{tabal13} and Tab.~\ref{tabal22} show the best-fit values of the parameters in {\sc relxill\_nk}, the reflection fraction, and the quality of the fit with $\chi^2$ and the number of degrees of freedom $\nu$. The reflection fraction, $R_{\rm ref}$, is calculated as the ratio between the flux of the relativistic reflection component and the flux of the power law and the relativistic reflection components in the energy range 0.6-10~keV.

In what follows we briefly describe the fit of every source and the measurements of the metric parameters $a_*$, $\alpha_{13}$, and $\alpha_{22}$.

\subsection{Ton~S180}

For this source we obtain a good fit with a relativistic reflection spectrum and a power law continuum. There is no detection of absorption or emission features. The XSPEC model is thus
\be
\text{\sc tbabs*(zpowerlw + relxill\_nk)} \, . \nonumber
\ee
The best-fit values of the {\sc relxill\_nk} parameters are reported in Tab.~\ref{tabal13} ($\alpha_{13}$ free and $\alpha_{22}=0$) and Tab.~\ref{tabal22} ($\alpha_{13}=0$ and $\alpha_{22}$ free). The high spin measurement and the high inner emissivity index are consistent with previous studies in which the Kerr metric was assumed~\cite{w13,nard12}. Fig.~\ref{f-180} shows the constraints on the spin parameter $a_*$ and on the deformation parameter $\alpha_{13}$ (left panel) and $\alpha_{22}$ (right panel). The red, green, and blue curves correspond, respectively, to the 68\%, 90\%, and 99\% confidence level boundaries for two relevant parameters. For $\alpha_{22}=0$, the constraints on spin and $\alpha_{13}$ are (here and in what follows we always report the 99\% confidence level for two parameters of interest)
\be
a_* > 0.987 \, , \qquad
-0.5 < \alpha_{13} < 0.1 \, .
\ee
For $\alpha_{13} = 0$, we find
\be
a_* > 0.984 \, , \qquad
-0.1 < \alpha_{22} < 0.9 \, .
\ee
When the value of the deformation parameter is consistent with 0, the measurement confirms the Kerr nature of the black hole. We note that we find an extremely high reflection fraction ($R_{\rm ref} > 0.99$) and the primary power law is orders of magnitudes below in flux (in Fig.~\ref{r-a13} and Fig.~\ref{r-a22} the black and red curves overlap for this source and we only see the black one).


\subsection{RBS~1124}

For this source we attain a good fit with a relativistic reflection component and a power law continuum. The XSPEC model is thus
\be
\text{\sc tbabs*(zpowerlw + relxill\_nk)} \, . \nonumber
\ee
The best-fit values are reported in Tab.~\ref{tabal13} ($\alpha_{13}$ free and $\alpha_{22}=0$) and Tab.~~\ref{tabal22} ($\alpha_{13}=0$ and $\alpha_{22}$ free). The constraints on the spin parameter and the deformation parameters are shown in Fig.~\ref{f-1124}. The gray regions are ignored in our analysis because they violate the constraints in Eq.~(\ref{eq-constraints}). As we can see from the two panels in Fig.~\ref{f-1124}, there are several local minima and thus several measurements. The constraints on $\alpha_{13}$ and $\alpha_{22}$ are eventually quite weak.


\subsection{Ark~120}

For this source, a power law continuum with a relativistic reflection component are not sufficient to describe the spectrum. The residuals show a narrow emission line around 6.4~keV and we add {\sc xillver} to describe a cold distant reflector. We also observe an emission feature around 6.9~keV and an absorption 
feature around 6~keV and we add two narrow gaussian lines. The emission feature is measured at the energy $6.95\pm0.03$~keV, which is consistent with Fe~XXVI. The absorption feature is found at the energy $6.087\pm0.015$~keV. The total XSPEC model is
\be
\text{\sc tbabs*(zpowerlw + relxill\_nk + xillver} \nonumber\\
\text{\sc  + zgauss + zgauss)} \, . \nonumber 
\ee
The constraints on $a_*$, $\alpha_{13}$, and $\alpha_{22}$ are shown in Fig.~\ref{f-120}. Tab.~\ref{tabal13} and Tab.~~\ref{tabal22} show the best-fit values in {\sc relxill\_nk}. Our estimate of the spin is higher than what was found in~\cite{nard11,patr11}, but we use a different reflection model and it is well know that with {\sc xillver} we recover higher spins. In the case $\alpha_{22}=0$, the constraints on spin $a_*$ and $\alpha_{13}$ are
\be
a_* > 0.991 \, , \qquad
-0.36 < \alpha_{13} < 0.08 \, .
\ee
When we assume $\alpha_{13} = 0$, the constraints on spin $a_*$ and $\alpha_{22}$ are
\be
a_* > 0.989 \, , \qquad
-0.06 < \alpha_{22} < 0.34 \, .
\ee


\subsection{Swift~J0501.9--3239}

A model with a power law continuum with a relativistic reflection component still shows an additional emission line around 6.4~keV. In order to model this feature, we include a cold distant reflector. The total model is 
\be
\text{\sc tbabs*(zpowerlw + relxill\_nk + xillver)} \, . \nonumber 
\ee
We obtain already a good fit with an emissivity profile described by a simple power law. The best-fit values of the reflection component are reported in Tab.~\ref{tabal13} and Tab.~\ref{tabal22}. For $\alpha_{22}=0$, the constraints on spin $a_*$ and $\alpha_{13}$ are
\be
a_* > 0.979 \, , \qquad
-0.7 < \alpha_{13} < 0.1 \, .
\ee
For $\alpha_{13} = 0$, the constraints on spin $a_*$ and $\alpha_{22}$ are
\be
a_* > 0.971 \, , \qquad
-0.1 < \alpha_{22} < 0.8 \, .
\ee
The constraints $a_*$ vs $\alpha_{13}$ and $a_*$ vs $\alpha_{22}$ are shown in Fig.~\ref{f-swift}.


\subsection{1H0419--577}

As for the previous source, here we have to add a cold distant reflector to the power law continuum and relativistic reflection component. The total XSPEC model is thus
\be
\text{\sc tbabs*(zpowerlw + relxill\_nk + xillver)} \, . \nonumber 
\ee
A good fit is reached assuming an emissivity profile described by a simple power law. The spin measurement is extremely high and consistent with previous analysis in the Kerr metric~\cite{w10,f05}. A simple power law is enough to describe the emissivity profile and to get a good fit; see Tab.~\ref{tabal13} and Tab.~\ref{tabal22} for the best-fit values. The constraints on $a_*$, $\alpha_{13}$, and $\alpha_{22}$ are shown in Fig.~\ref{f-0419}. For $\alpha_{22}=0$, the estimates of $a_*$ and $\alpha_{13}$ are
\be
a_* > 0.988 \, , \qquad
-0.35 < \alpha_{13} < 0.12 \, .
\ee
For $\alpha_{13} = 0$, we obtain 
\be
a_* > 0.987 \, , \qquad
-0.08 < \alpha_{22} < 0.32 \, .
\ee


\subsection{PKS~0558--504}

In addition to the power law and relativistic reflection components, we need a narrow emission line around 7~keV. For the latter we use {\sc zgauss}, and the fit finds the energy $6.95\pm0.05$~keV, which corresponds to Fe~XXVI. The total model reads
\be
\text{\sc tbabs*(zpowerlw + relxill\_nk + zgauss)} \, . \nonumber 
\ee
The contours are shown in Fig.~\ref{f-504}. The best-fit values are reported in Tab.~\ref{tabal13} and Tab.~\ref{tabal22}. The high spin value is in agreement with previous studies~\cite{g10,nard11,w13}. Here the inclination angle is free and we find a moderate value, while in Ref.~\cite{w13} the authors find a very high value of the inclination angle inconsistent with this source and therefore freeze it to $45^\circ$. For $\alpha_{22}=0$, the constraints on $a_*$ and $\alpha_{13}$ are
\be
a_* > 0.989 \, , \qquad
-0.4 < \alpha_{13} < 0.1 \, .
\ee
For $\alpha_{13} = 0$, the constraints on $a_*$ and $\alpha_{22}$ are 
\be
a_* > 0.987 \, , \qquad
-0.1 < \alpha_{22} < 0.5 \, .
\ee


\subsection{Fairall~9}

After fitting the spectrum with the power law and relativistic reflection components, we see two narrow emission lines. We add a non-relativistic reflection component and a gaussian line. The latter is found to be at $6.97\pm0.04$~keV and can be interpreted as a Fe~XXVI line~\cite{w13}. The total model is
\be
\text{\sc tbabs*(zpowerlw + relxill\_nk} \nonumber\\
\text{\sc  + xillver + zgauss)} \, . \nonumber 
\ee
Fig.~\ref{f-9} shows the constraints on $a_*$, $\alpha_{13}$, and $\alpha_{22}$. Tab.~\ref{tabal13} and Tab.~\ref{tabal22} show the best-fit values of {\sc relxill\_nk}. For $\alpha_{22}=0$, we find
\be
a_* > 0.92 \, , \qquad
-1.0 < \alpha_{13} < 0.3 \, .
\ee
For $\alpha_{13} = 0$, the constraints are 
\be
a_* > 0.91 \, , \qquad
0.0 < \alpha_{22} < 2.1 \, .
\ee

\begin{figure*}[t]
\begin{center}
\includegraphics[width=8.5cm]{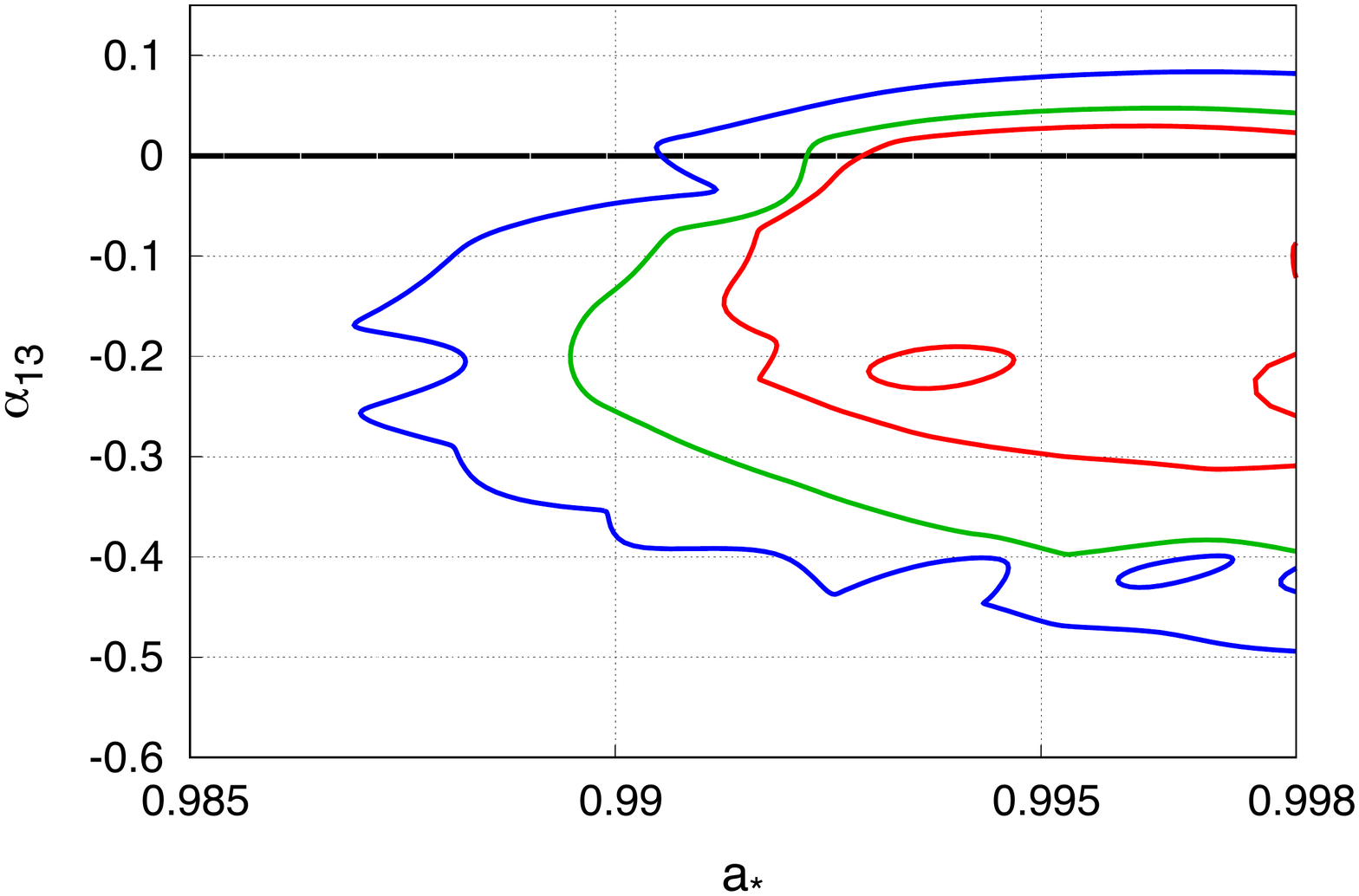}
\includegraphics[width=8.5cm]{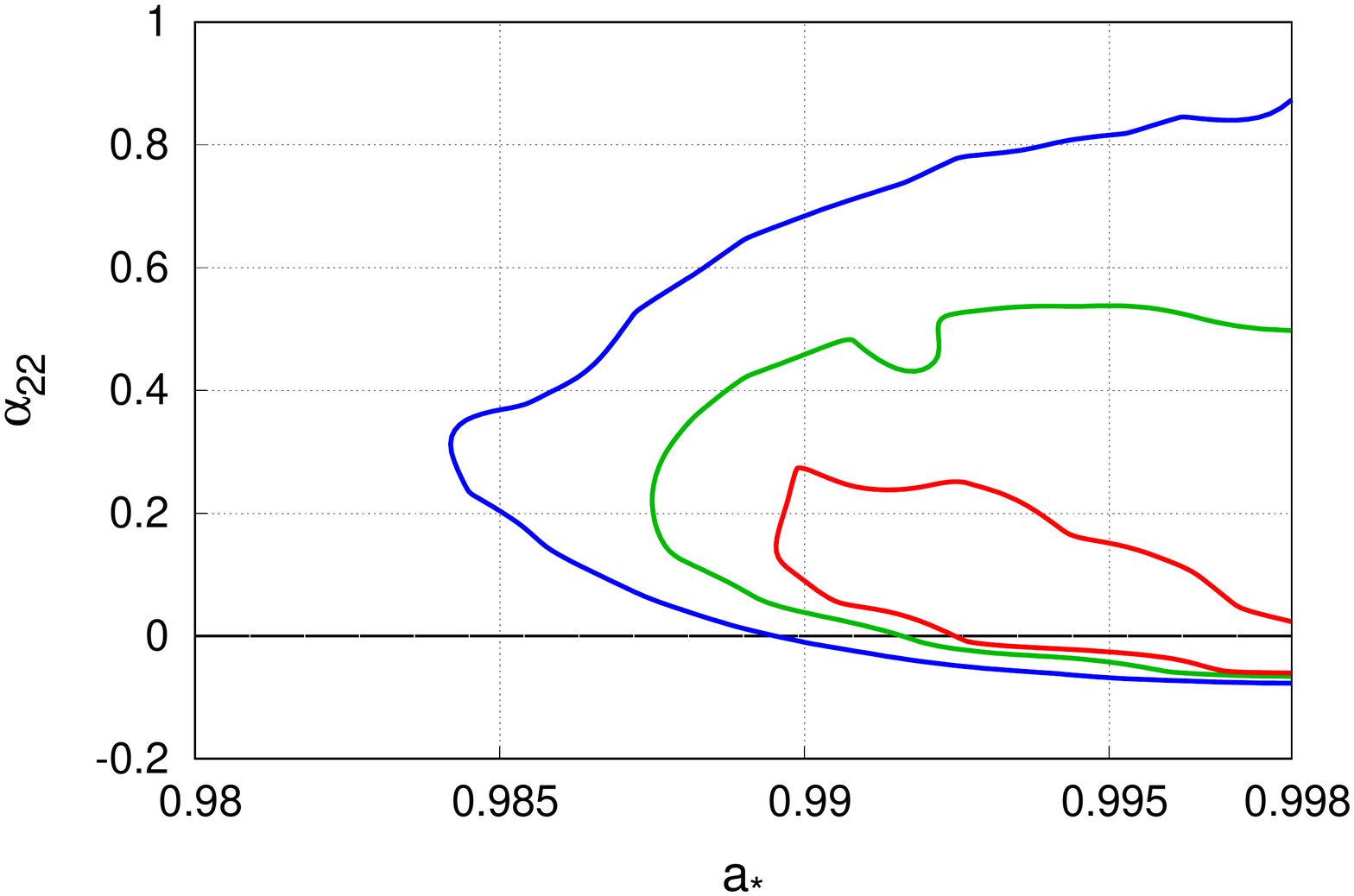}
\end{center}
\vspace{-1.2cm}
\caption{Constraints on the spin parameter $a_*$ and the Johannsen deformation parameters $\alpha_{13}$ (left panel) and $\alpha_{22}$ (right panel) for the supermassive black hole in Ton~S180. The red, green, and blue curves are, respectively, the 68\%, 90\%, and 99\% confidence level boundaries for two relevant parameters. \label{f-180}}
\begin{center}
\includegraphics[width=8.5cm]{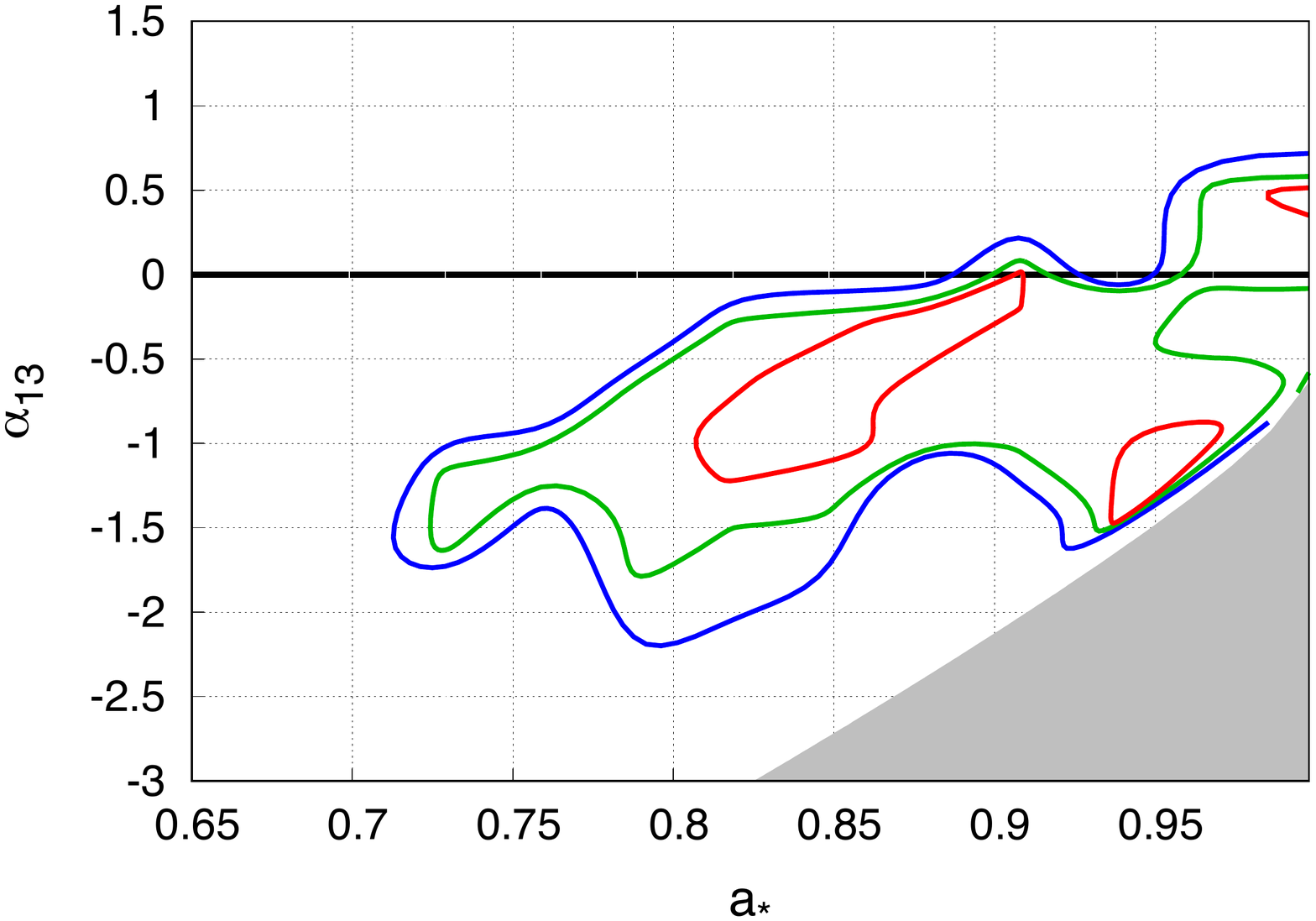}
\includegraphics[width=8.5cm]{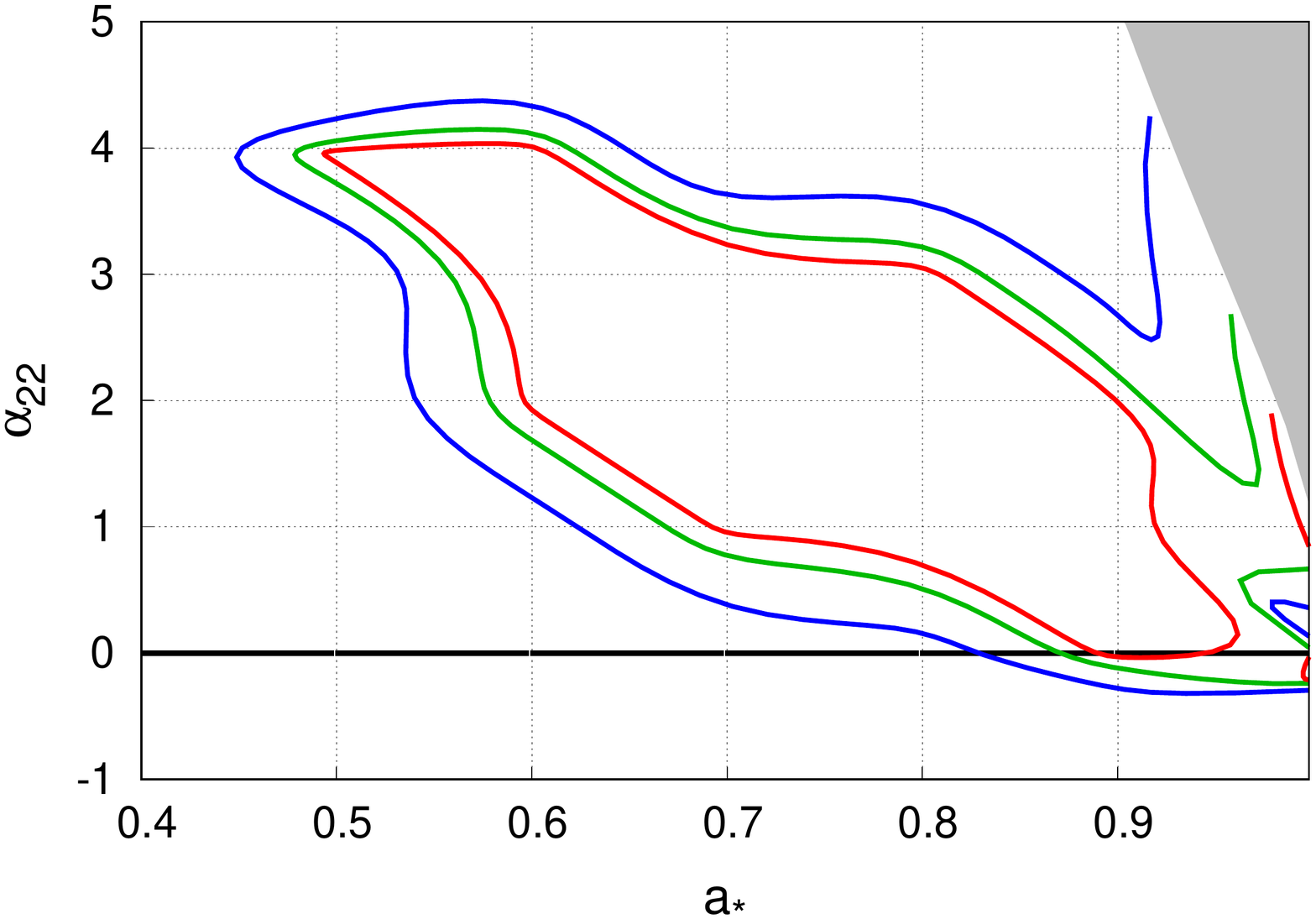}
\end{center}
\vspace{-1.2cm}
\caption{As in Fig.~\ref{f-180} for the supermassive black hole in RBS~1124. \label{f-1124}}
\begin{center}
\includegraphics[width=8.5cm]{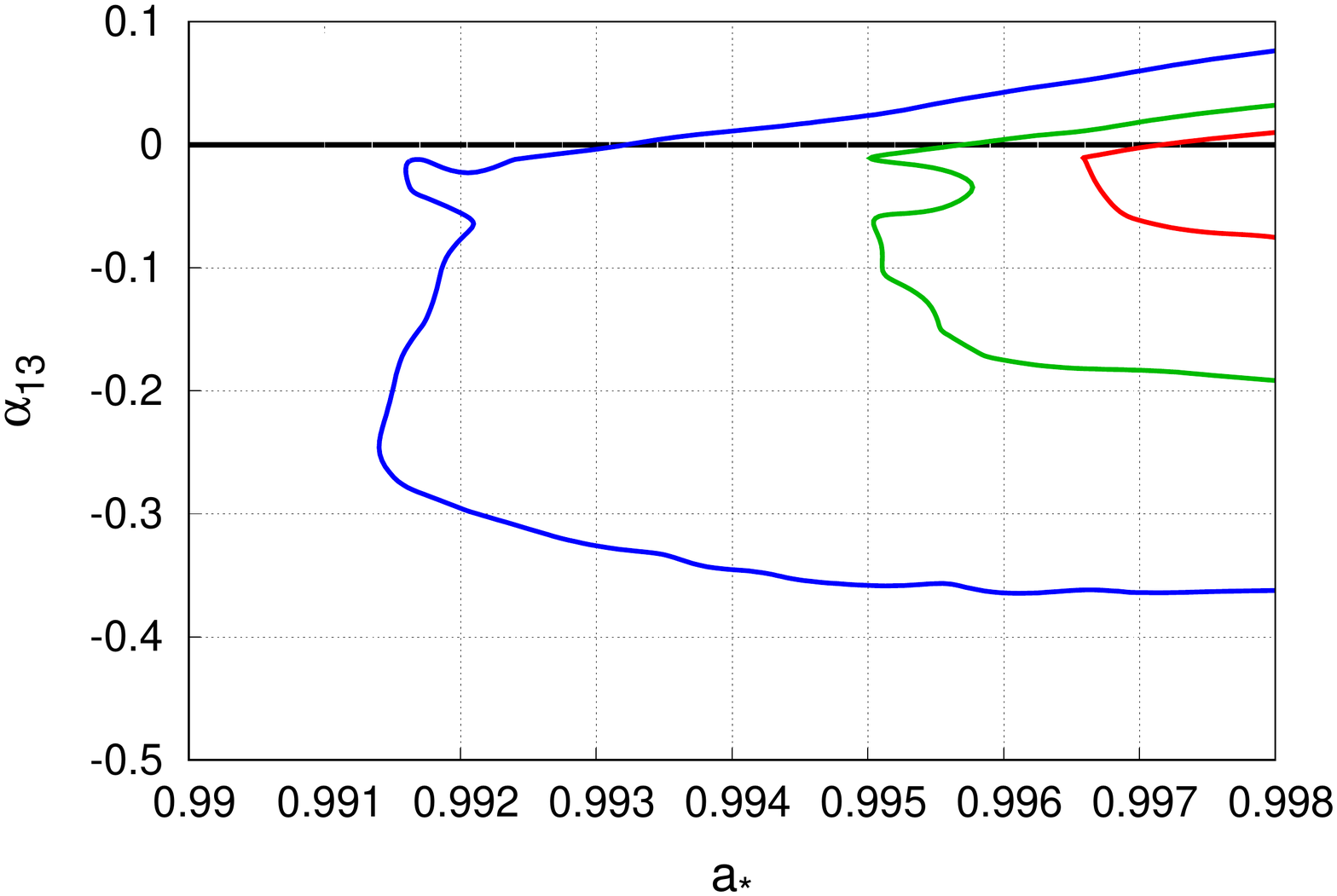}
\includegraphics[width=8.5cm]{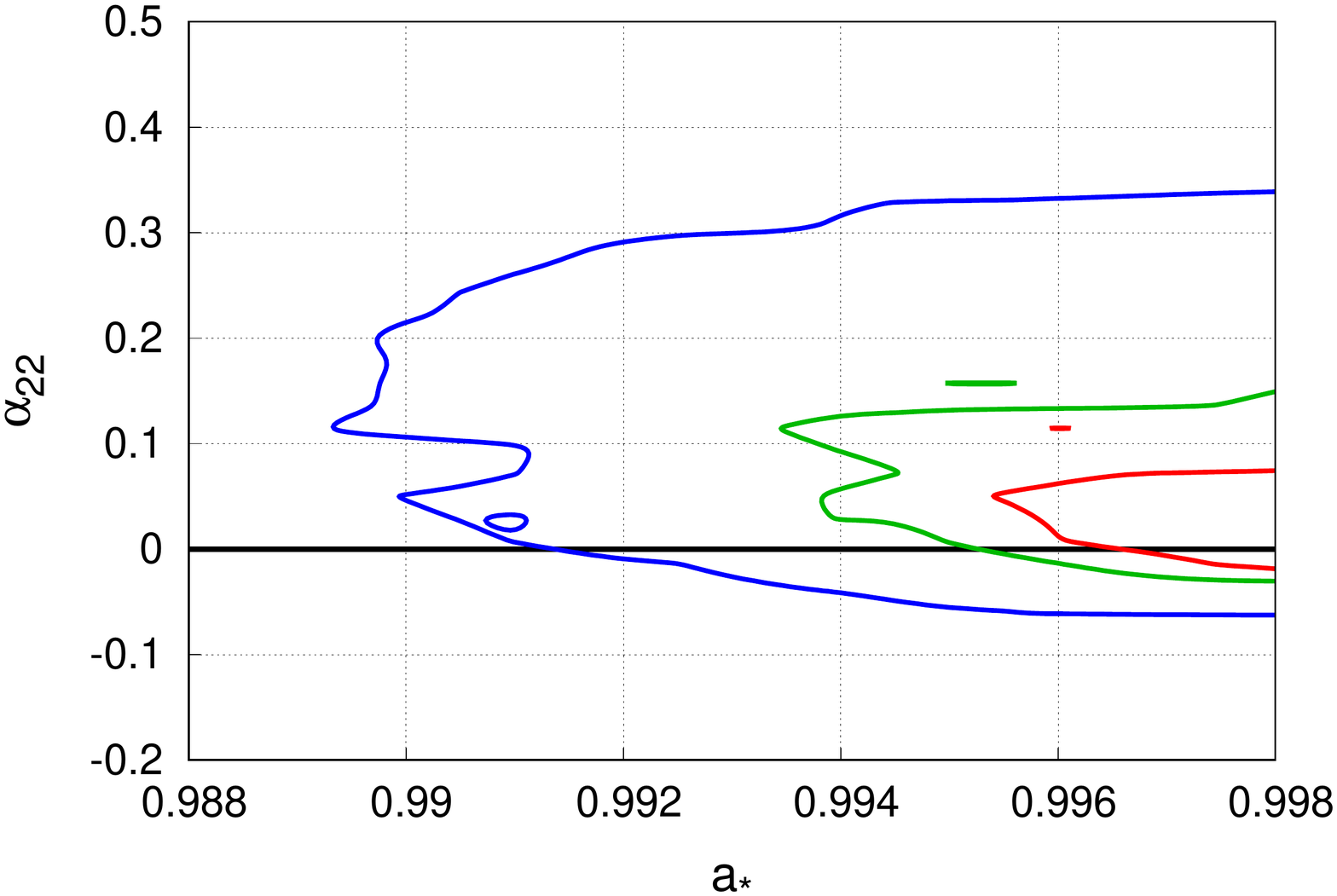}
\end{center}
\vspace{-1.2cm}
\caption{As in Fig.~\ref{f-180} for the supermassive black hole in Ark~120.} \label{f-120}
\end{figure*}

\begin{figure*}[t]
\begin{center}
\includegraphics[width=8.5cm]{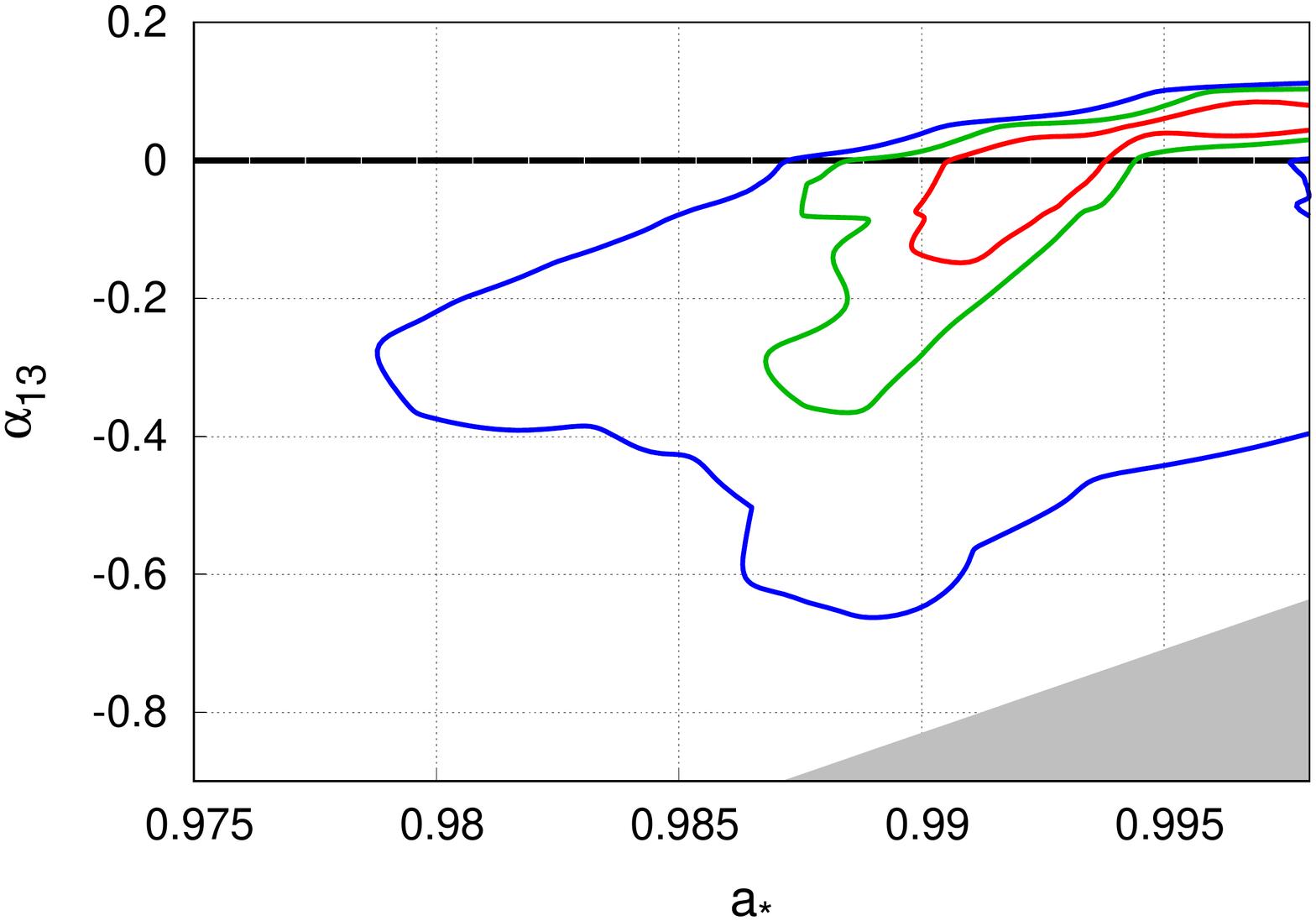}
\includegraphics[width=8.5cm]{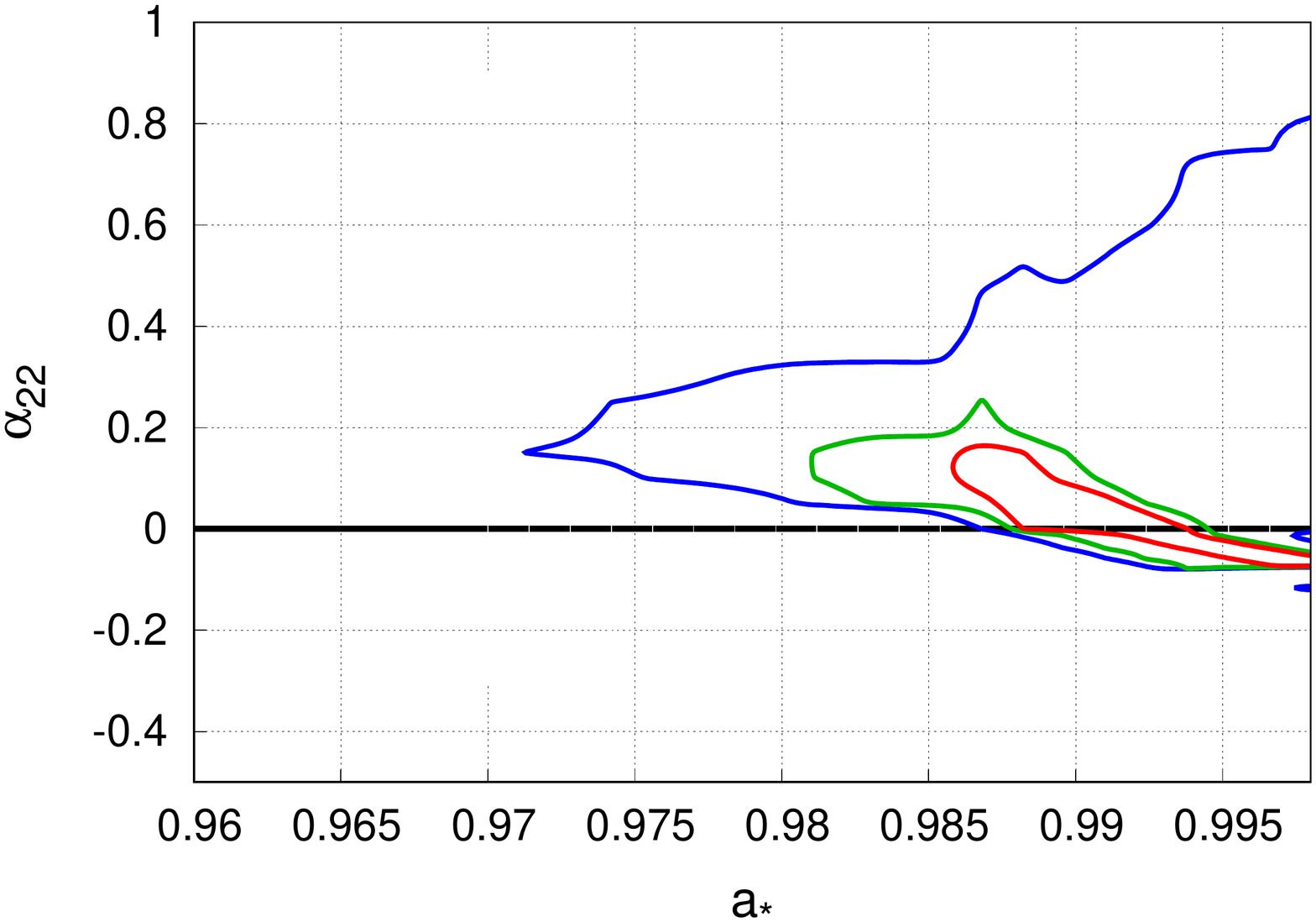}
\end{center}
\vspace{-1.2cm}
\caption{As in Fig.~\ref{f-180} for the supermassive black hole in Swift~J0501.9--3239. \label{f-swift}}
\begin{center}
\includegraphics[width=8.5cm]{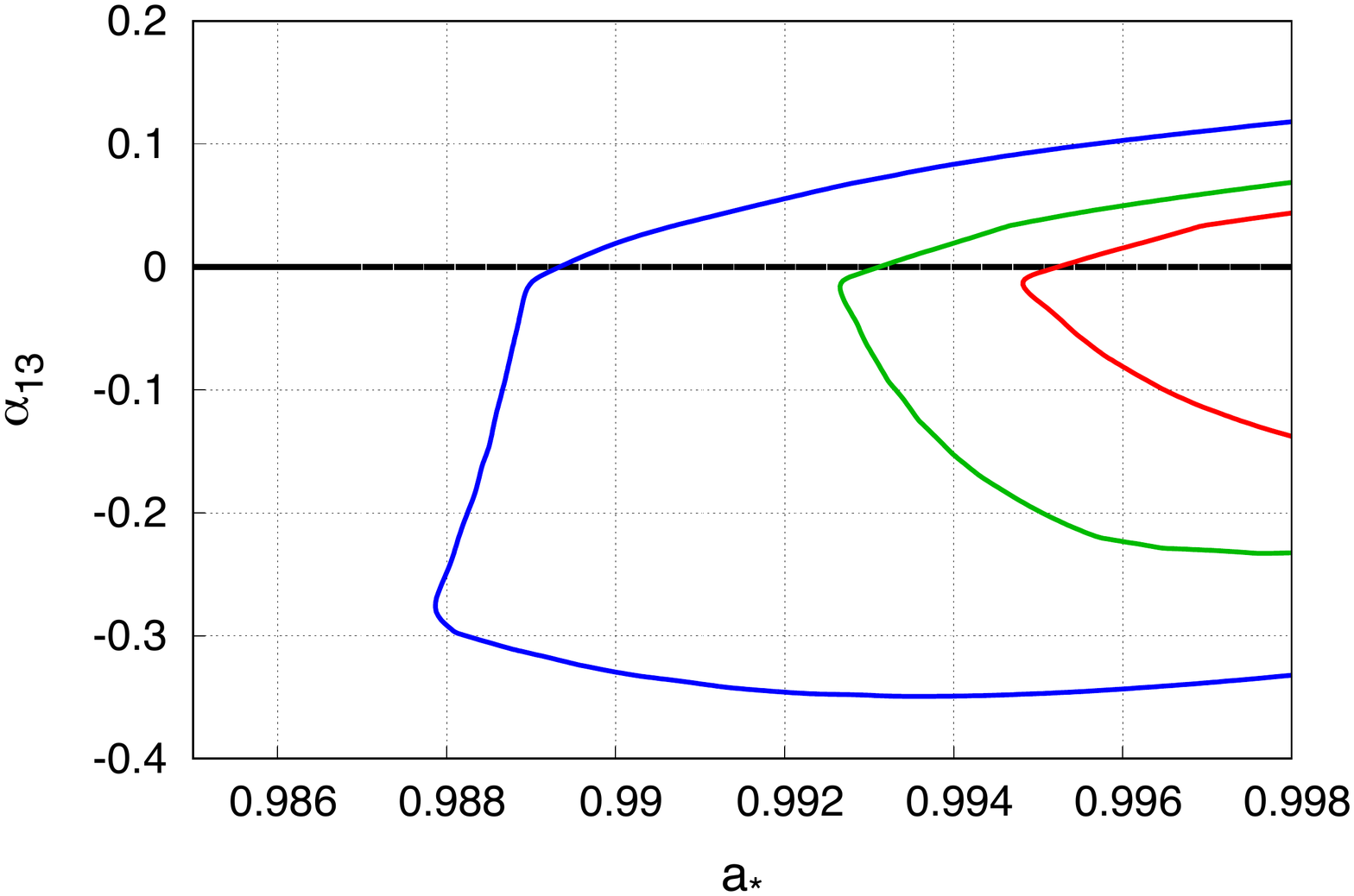}
\includegraphics[width=8.5cm]{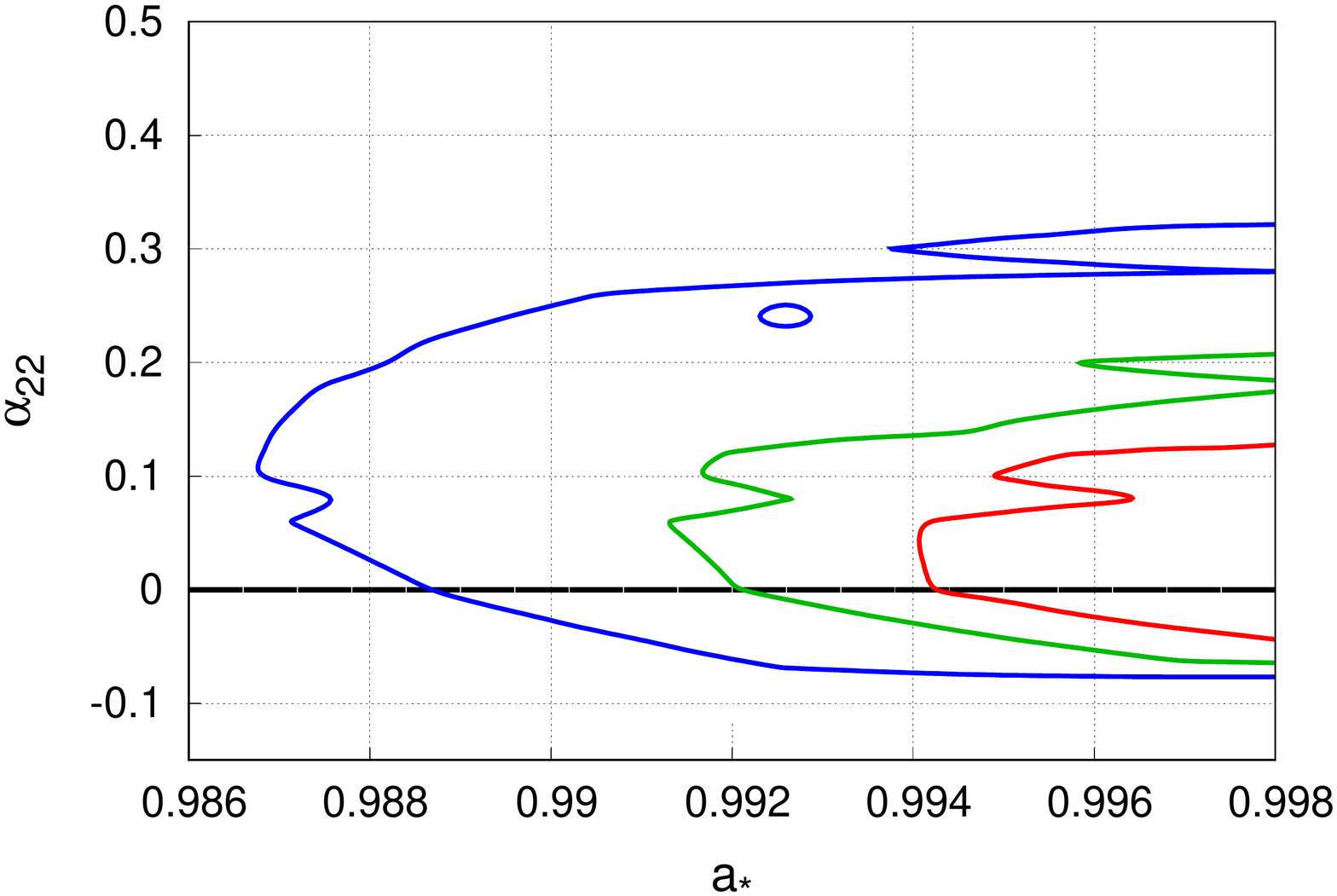}
\end{center}
\vspace{-1.2cm}
\caption{As in Fig.~\ref{f-180} for the supermassive black hole in 1H0419--577. \label{f-0419}}
\end{figure*}

\begin{figure*}[t]
\begin{center}
\includegraphics[width=8.5cm]{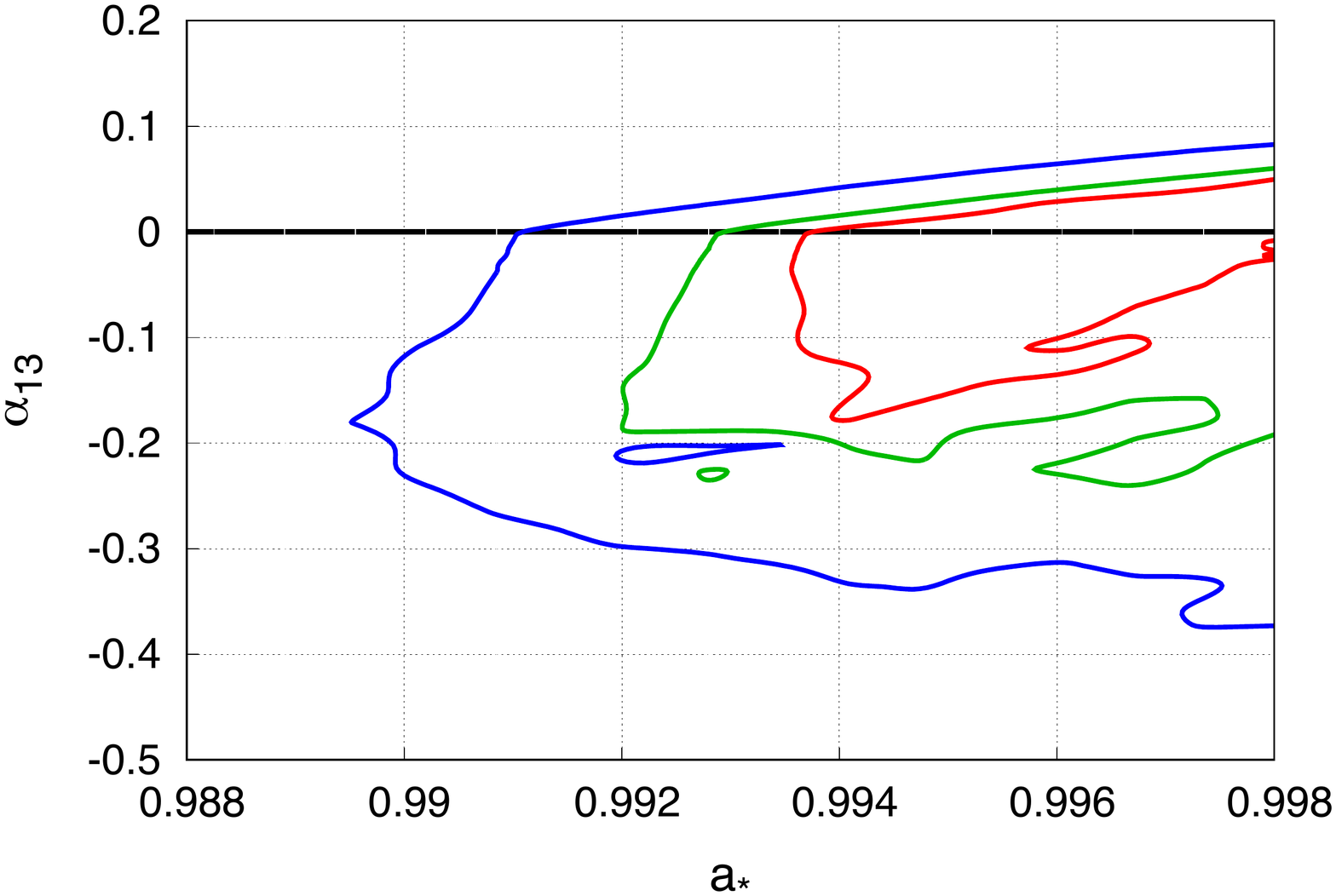}
\includegraphics[width=8.5cm]{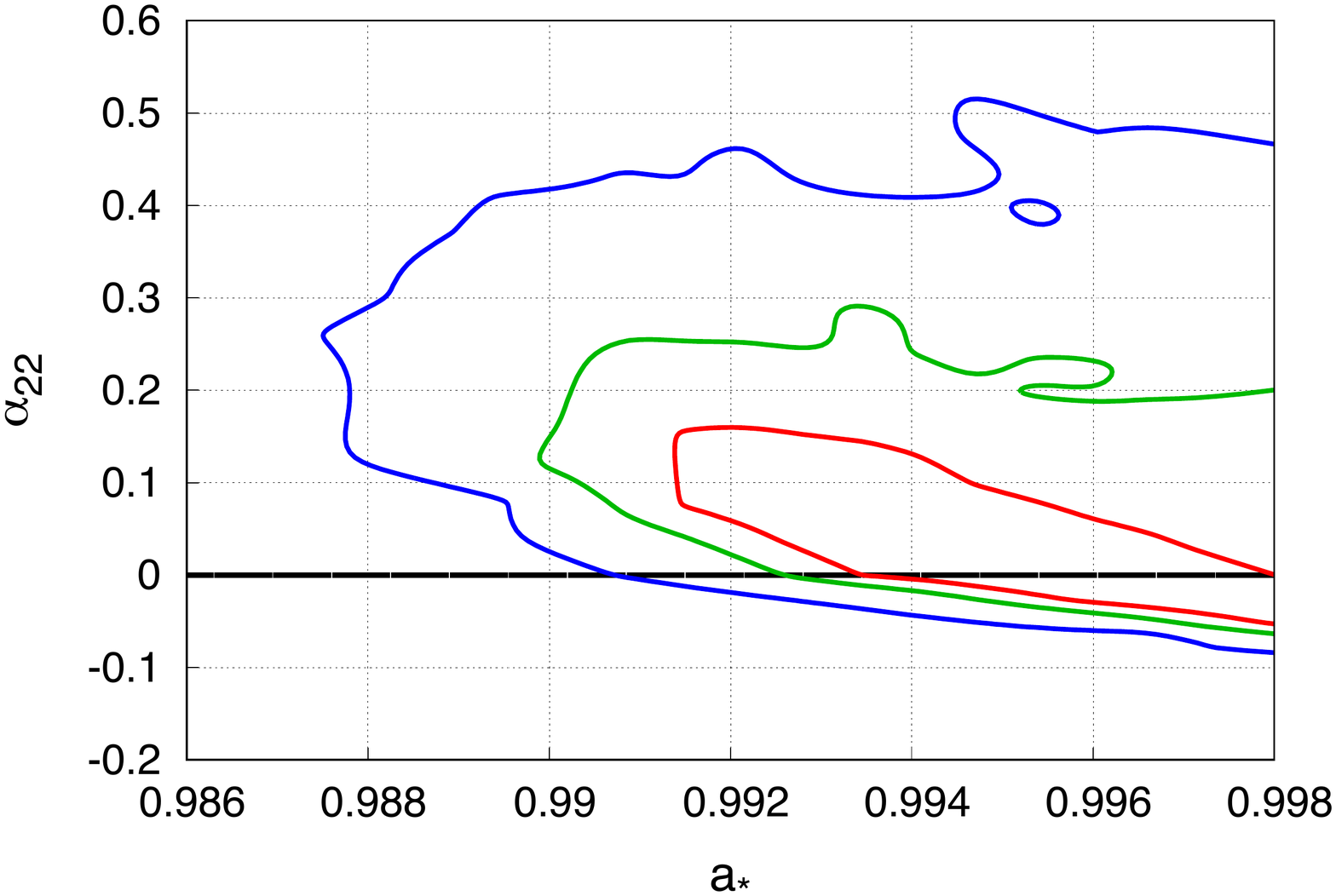}
\end{center}
\vspace{-1.2cm}
\caption{As in Fig.~\ref{f-180} for the supermassive black hole in PKS~0558--504. \label{f-504}}
\begin{center}
\includegraphics[width=8.5cm]{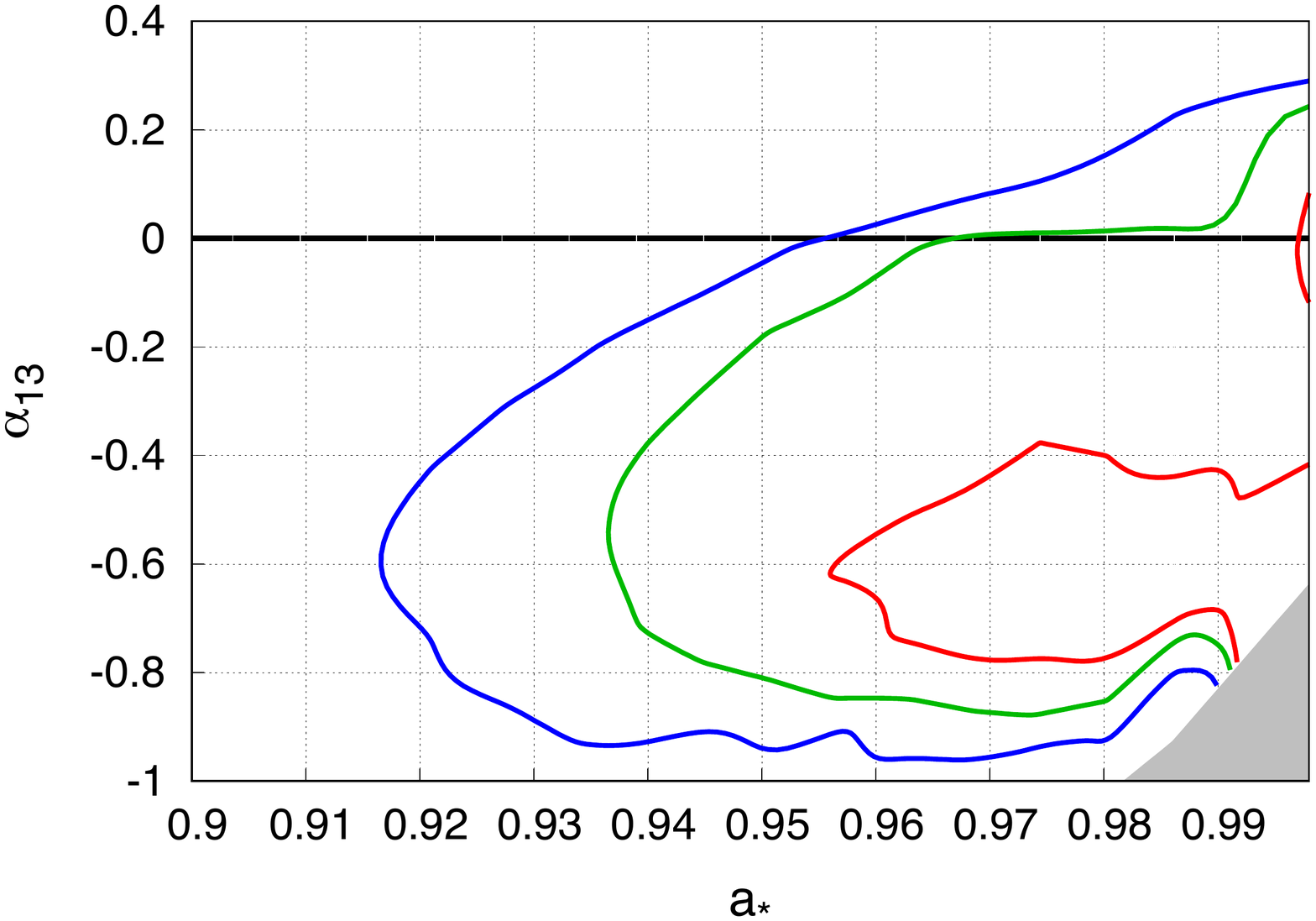}
\includegraphics[width=8.5cm]{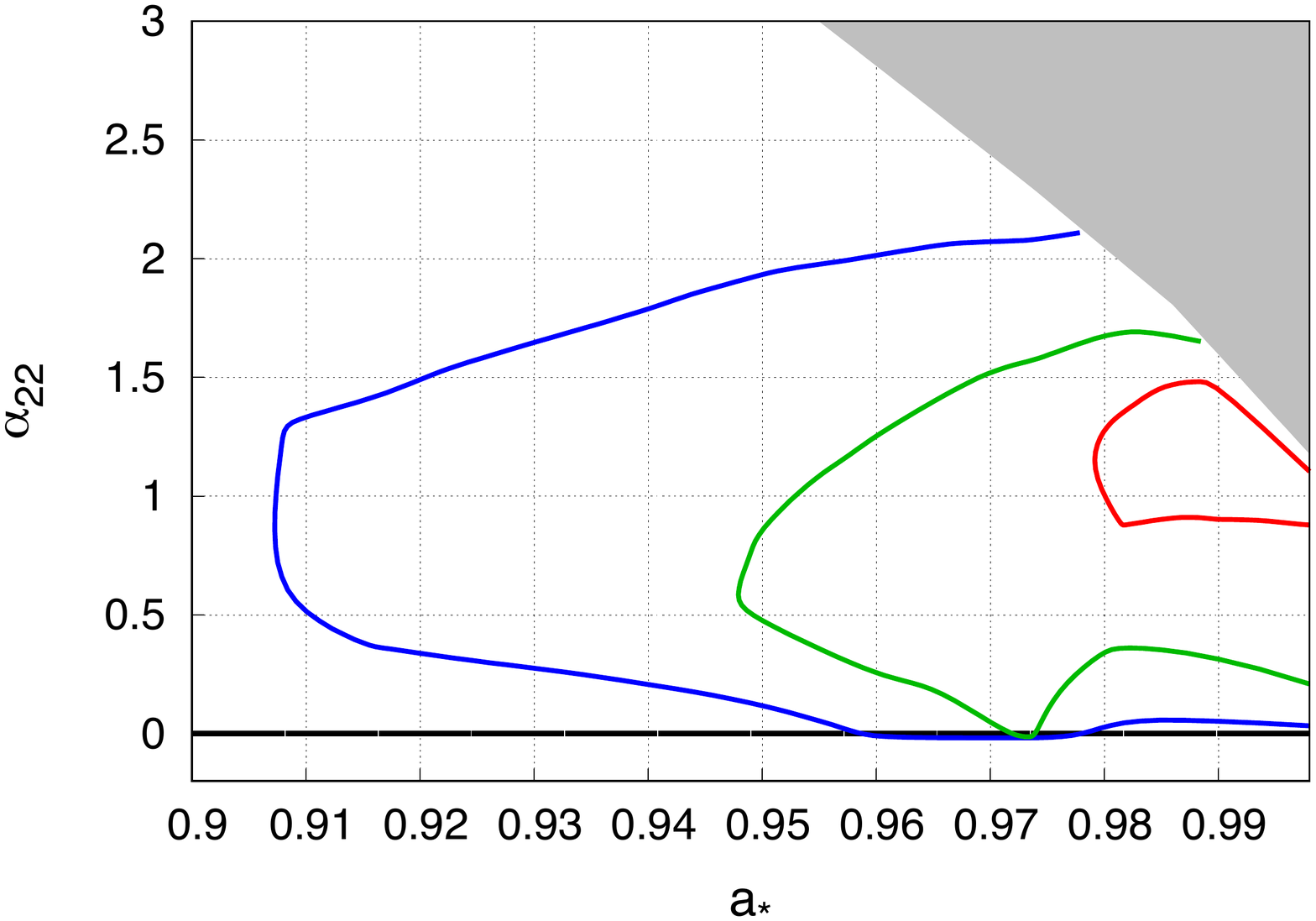}
\end{center}
\vspace{-1.2cm}
\caption{As in Fig.~\ref{f-180} for the supermassive black hole in Fairall~9. \label{f-9}}
\end{figure*}


\section{Discussion and conclusions \label{s-dis}}

In the previous section, we have presented our analysis of seven bare AGN observed with \textsl{Suzaku}. All these sources have a simple spectrum, with little or no intrinsic absorption, and are therefore good candidates for our study aimed at testing general relativity in the strong field regime. Interpreting the spectra of these sources as dominated by the power law from the corona and by the disk reflection spectrum, we have employed our relativistic reflection model {\sc relxill\_nk} and constrained the Johannsen deformation parameters $\alpha_{13}$ and $\alpha_{22}$.

Our results are largely consistent with previous studies in which the Kerr metric is assumed, in particular the analysis reported in Ref.~\cite{w13}. In some cases, our measurements of the spin parameters are somewhat higher and those of the inclination angle somewhat lower, but this can be attributed to the different non-relativistic reflection model. The study in~\cite{w13} employs the non-relativistic reflection model {\sc reflionx}~\cite{ross05}, here we have {\sc xillver}.

Some sources do not show a prominent blurred iron line around 6~keV. Strong constraints on the black hole spin and the Johannsen deformation parameters are possible from the high photon count in the soft X-ray band ($< 1$~keV): interpreting the spectra as reflection dominated leads to measurement of very high values of the spin parameter and deformation parameters close to 0. For example, this is the case of Ark~120 and Fairall~9. RBS~1124 does not show an iron line at all: the constraints on $\alpha_{13}$ and $\alpha_{22}$ are weak, but they are still possible from the excess of counts at low energies.

We model the emissivity profile of the accretion disk with a power law or a broken power law, and we always find a high value of the emissivity index or of the inner emissivity index, respectively. This indicates that most of the emission comes from the very inner region of the disk, and is therefore strongly affected by relativistic effects. Very high emissivity indices were also found in Ref.~\cite{w13}.

\begin{figure*}[t]
\begin{center}
\includegraphics[width=8.5cm]{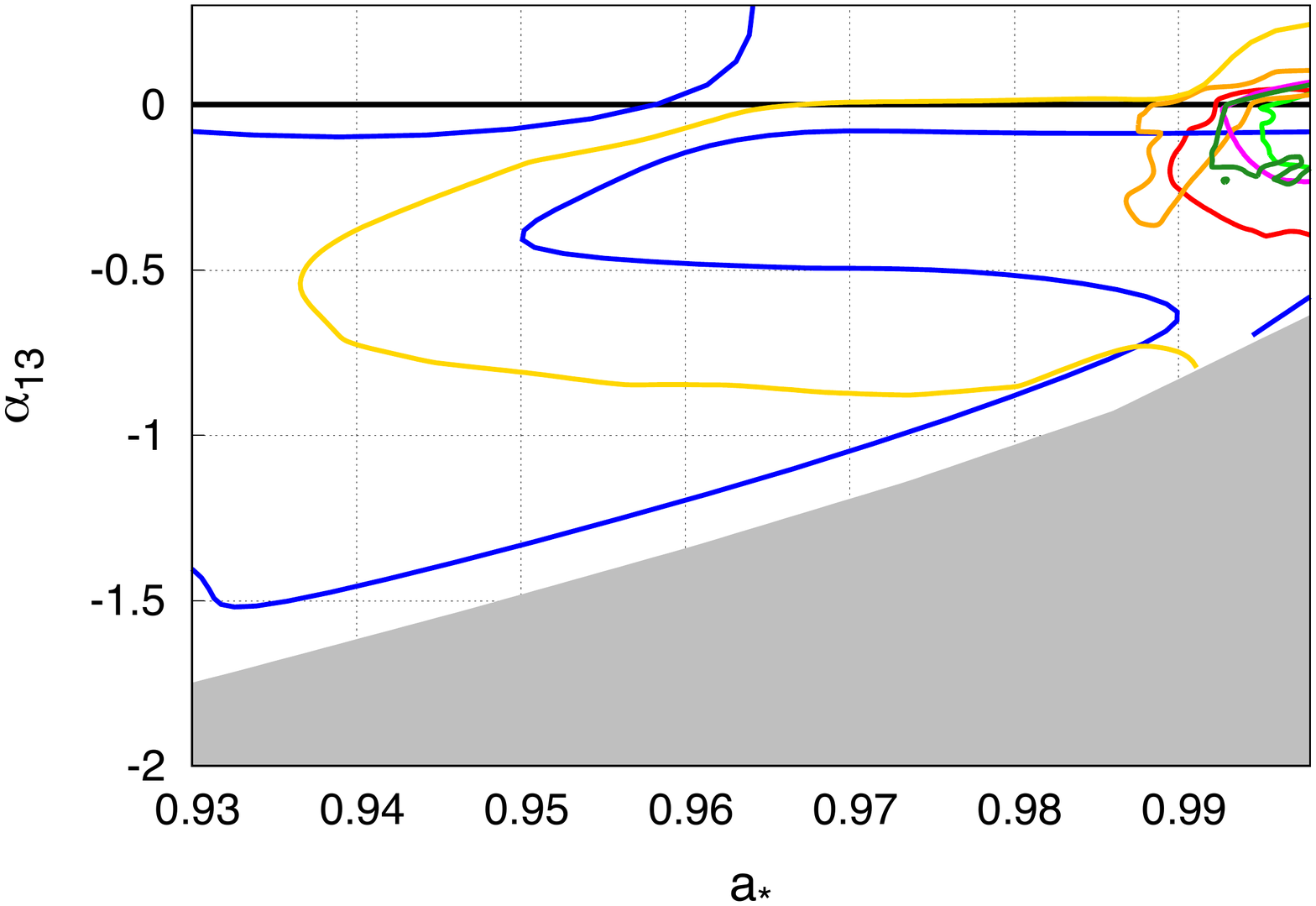}
\includegraphics[width=8.5cm]{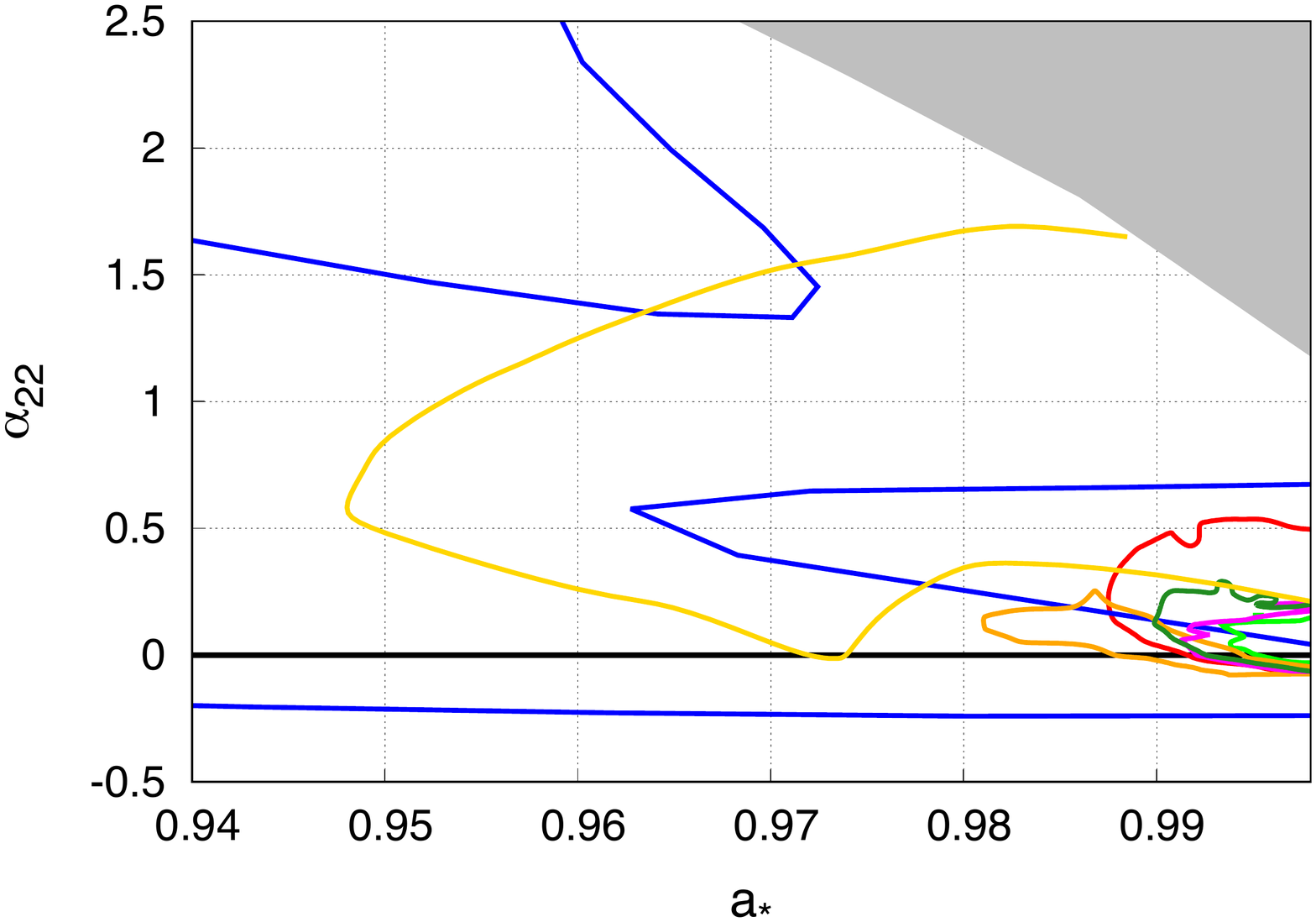}
\end{center}
\vspace{-1.2cm}
\caption{90\% confidence level constraints on the spin parameter $a_*$ and the Johannsen deformation parameters $\alpha_{13}$ (left panel) and $\alpha_{22}$ (right panel) of the seven sources studied in this paper: Ton~S180 (red), RBS~1124 (blue), Ark~120 (light green), Swift~J0501.9--3239 (orange), 1H0419--577 (magenta), PKS~0558--504 (forest green), and Fairall~9 (yellow). \label{f-summary}}
\end{figure*}

All our results are consistent with the hypothesis that the spacetime metric around these sources is described by the Kerr solution of general relativity. The constraints from Ton~S180, Ark~120, Swift~J0501.9--3239, 1H0419--577, and PKS~0558--504 are quite stringent and comparable to the constraints inferred from GS~1354--645 in Ref.~\cite{noi4} 
\be
&& a_* > 0.975 \, , \,\,
-0.34 < \alpha_{13} < 0.16 \quad (\text{for } \alpha_{22} = 0) \, , \quad \nonumber\\
&& a_* > 0.975 \, , \,\,
-0.09 < \alpha_{22} < 0.42 \quad (\text{for } \alpha_{13} = 0) \, , \quad \nonumber
\ee
and from MCG--6--30--15 in Ref.~\cite{noi6}
\be
&& 0.928 < a_* < 0.983 \, , \,\,
-0.44 < \alpha_{13} < 0.15 \quad (\text{for } \alpha_{22} = 0) \, , \quad \nonumber\\
&& 0.885 < a_* < 0.987 \, , \,\,
-0.12 < \alpha_{22} < 1.05 \quad (\text{for } \alpha_{13} = 0) \, . \quad \nonumber
\ee
This may be initially surprising, because the quality of the data analyzed in Refs.~\cite{noi4,noi6} is better, as GS~1354--645 is a stellar-mass black hole and MCG--6--30--15 is a very bright AGN with observations of both \textsl{XMM-Newton} and \textsl{NuSTAR}. The key points are likely $i)$ the very high spins of the sources in this paper and $ii)$ their simple spectra. The former allows for a better probe of the strong gravity region, the latter limits the parameter degeneracy.
Fig.~\ref{f-summary} shows the 90\% confidence level constraints on the spin parameter $a_*$ and the Johannsen deformation parameters $\alpha_{13}$ and $\alpha_{22}$ of the seven sources together (see the figure caption for the color associated to every source).

A weak point in our analysis is represented by our disk model. {\sc relxill\_nk} assumes that the disk is thin and the inner edge is at the ISCO radius. Moreover, the thickness of the disk is completely ignored, and the disk is assumed to be infinitesimally thin. Thin disks with the inner edge at the ISCO radius can be expected for mass accretion rates roughly between 5\% to 30\% of the Eddington limit, and the thickness of the disk increases as the mass accretion rate increases. The mass accretion rate in AGN is usually difficult to estimate, because of the poor estimates of the distance, mass, and total accretion luminosity. However, most sources are thought to accrete above the 30\% Eddington limit. In such a case, the disk is likely slim or thick, and the inner edge may be inside the ISCO radius. This is often thought to lead to overestimates of the spin parameter of the black hole.

\begin{figure*}[t]
\begin{center}
\includegraphics[width=8.5cm]{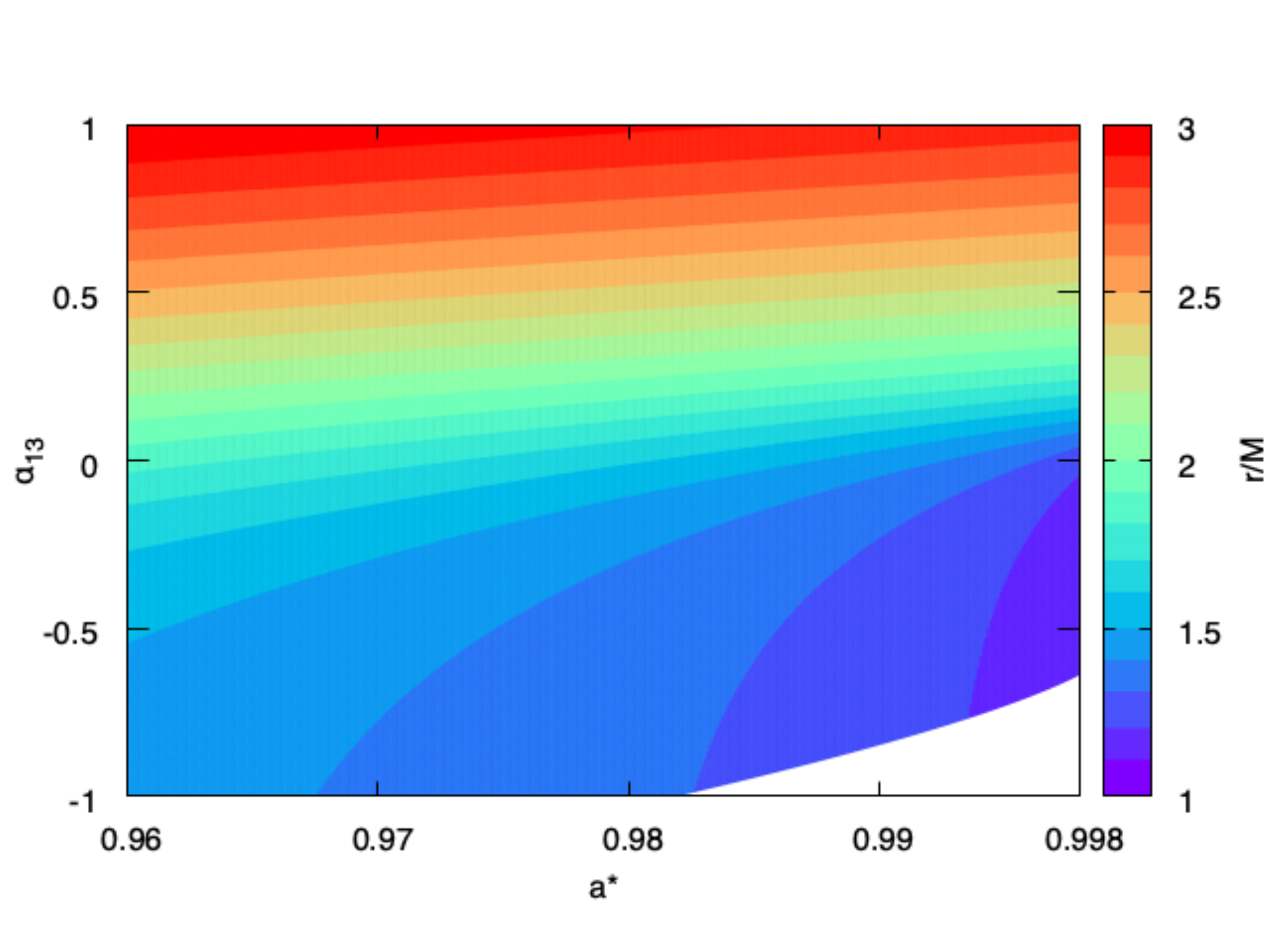}
\includegraphics[width=8.5cm]{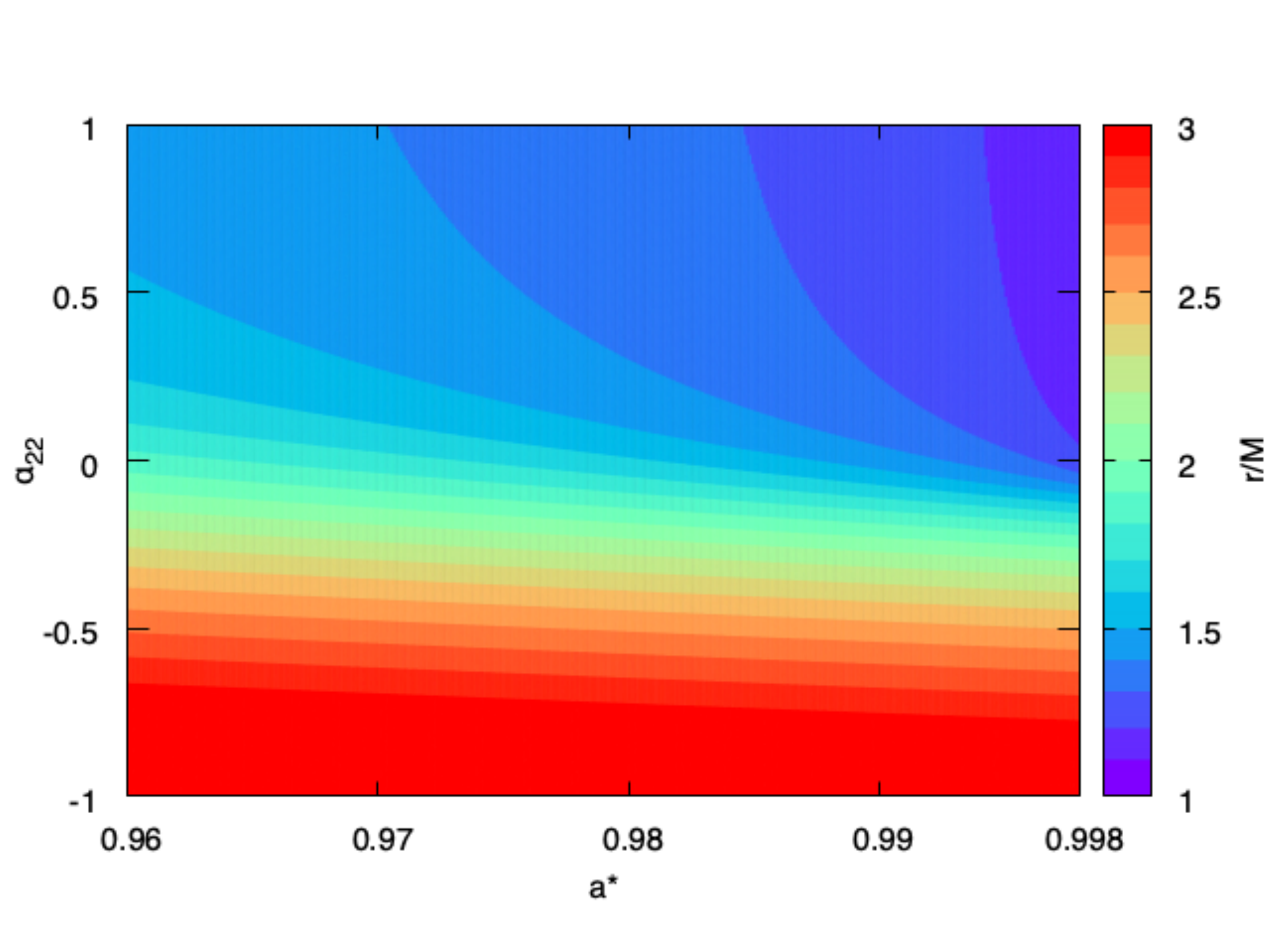}
\end{center}
\vspace{-0.7cm}
\caption{Contour map of the ISCO radius in the plane $a_*$ vs $\alpha_{13}$ (left panel) and $a_*$ vs $\alpha_{22}$ (right panel). The white area in the left panel is ignored because it does not meet the conditions in Eq.~(\ref{eq-constraints}). \label{f-isco}}
\end{figure*}

We can thus question whether the strong constraints on $\alpha_{13}$ and $\alpha_{22}$ from Ton~S180, Ark~120, Swift~J0501.9--3239, 1H0419--577, and PKS~0558--504 can be simply attributed to the fact that the inner edge of the disk is very close to the compact object, probably inside the ISCO radius because of the high mass accretion rate. In other words, if the ISCO radius moves to very small radii only in the case of Kerr black holes with $a_* \rar 1$, any object with an inner edge of the disk very close to the object itself could be incorrectly interpreted as a Kerr black hole with very high spin and disk inner edge at the ISCO radius. The answer is no. Fig.~\ref{f-isco} shows the ISCO radius on the plane spin vs $\alpha_{13}$ (left panel) and spin vs $\alpha_{22}$ (right panel). As we can see from these plots and from the constraints from Ton~S180, Ark~120, Swift~J0501.9--3239, 1H0419--577, and PKS~0558--504, the confidence level curves of the measurements of $\alpha_{13}$ and $\alpha_{22}$ do not follow the contour level curves of the ISCO. Moreover, {\sc relxill\_nk} includes spins up to 0.998 and for negative values of $\alpha_{13}$ and positive values of $\alpha_{22}$ the ISCO radius can be smaller than the ISCO radius in the Kerr metric with $a_* = 0.998$.

Simplifications in the model are not limited to the disk morphology. The calculation of the non-relativistic reflection spectrum with {\sc xillver} assumes a constant disk density. Thus, the disk electron density is fixed and we have a single ionization parameter for the whole disk. The emissivity profile is modeled by a power law or a broken power law, but this is clearly an approximation and cannot be the correct emissivity profile, whatever the corona geometry.

Eventually, it is quite surprising that we can obtain some strong constraints on $\alpha_{13}$ and $\alpha_{22}$ and we always recover the Kerr solution\footnote{As shown in Tabs.~\ref{tabal13} and \ref{tabal22}, all our measurements of $\alpha_{13}$ and $\alpha_{22}$ are consistent with the Kerr solution at 90\% confidence level with the exception of those from RBS~1124 and Fairall~9. In the cases of RBS~1124 (for both $\alpha_{13}$ and $\alpha_{22}$) and of Fairtall~9 for $\alpha_{13}$, actually we have several measurements: as we can see from Figs.~\ref{f-1124} and \ref{f-9}, there are also measurements consistent with the Kerr solution at 90\% confidence level. In the case of Fairall~9 for $\alpha_{22}$, the agreement with the Kerr solution is more marginal, but still we can recover the Kerr metric as shown in Fig.~\ref{f-9}.}, despite the clear simplifications in the model and the corresponding systematic uncertainties not fully under control. The uncertainties reported in the previous section are only statistical, and the systematic ones are completely ignored. We would thus be tempted to argue that the systematic uncertainties due to the disk model are subdominant with respect to the statistical ones, as a perfect cancellation from very different errors would be highly unlikely. However, this issue can only be fully addressed with a specific study of the systematic uncertainties, which is not the scope of the present manuscript, and work is underway.

{\bf Acknowledgments --}
This work was supported by the National Natural Science Foundation of China (NSFC), Grant No.~U1531117, and Fudan University, Grant No.~IDH1512060. A.T. also acknowledges support from the China Scholarship Council (CSC), Grant No.~2016GXZR89. A.B.A. also acknowledges the support from the Shanghai Government Scholarship (SGS). J.A.G. acknowledges support from the Alexander von Humboldt Foundation. J.J. is supported by the Cambridge Trust and the Chinese Scholarship Council Joint Scholarship Programme (201604100032). S.N. acknowledges support from the Excellence Initiative at Eberhard-Karls Universit\"at T\"ubingen.


\end{document}